\documentclass[aps,pra,preprint,showpacs]{revtex4-1}

\usepackage{graphicx}
\usepackage{dcolumn}
\usepackage{bm}
\usepackage[load-configurations  =  version-1]{siunitx}
\usepackage{hyperref}

\begin{document}


\title{ESR measurements of phosphorus dimers in isotopically enriched $^{28}$Si silicon} 



\author{S. Shankar}
\email[]{shyam.shankar@yale.edu}
\altaffiliation[Now at ]{Department of Applied Physics, Yale University, New Haven, CT 06511.}

\author{A. M. Tyryshkin}

\author{S. A. Lyon}
\affiliation{Dept.\ of Electrical Engineering, Princeton University, Princeton, NJ 08544, USA}


\date{\today}

\begin{abstract}
Dopants in silicon have been studied for many decades using optical and electron spin resonance (ESR) spectroscopy. Recently, new features have been observed in the spectra of dopants in isotopically enriched $^{28}$Si since the reduced inhomogeneous linewidth in this material improves spectral resolution. With this in mind, we measured ESR on exchange coupled phosphorus dimers in $^{28}$Si and report two results. First, a new fine structure is observed in the ESR spectrum arising from state mixing by the hyperfine coupling to the $^{31}$P nuclei, which is enhanced when the exchange energy is comparable to the Zeeman energy. This fine structure enables us to spectroscopically address two separate dimer sub-ensembles, the first with exchange ($J$) coupling ranging from $2$ to $7$~GHz and the second with $J$ ranging from $6$ to $60$~GHz. Next, the average spin relaxation times, $T_1$ and $T_2$ of both dimer sub-ensembles were measured using pulsed ESR at $0.35$~T. Both $T_1$ and $T_2$ for transitions between triplet states of the dimers were found to be identical to the relaxation times of isolated phosphorus donors in $^{28}$Si, with $T_2 =$ \SI{4}{\ms} at \SI{1.7}{\kelvin} limited by spectral diffusion due to dipolar interactions with neighboring donor electron spins. This result, consistent with theoretical predictions, implies that an exchange coupling of $2$--$60$~GHz does not limit the dimer $T_1$ and $T_2$ in bulk Si at the \SI{10}{\ms} timescale.
\end{abstract} 


\maketitle 



%
%

%

\section{Introduction}
\label{sec:Introduction}

Natural silicon contains three stable isotopes: \SI{92.2}{\percent} of $^{28}$Si, \SI{4.7}{\percent} of $^{29}$Si, and \SI{3.1}{\percent} of $^{30}$Si. A concerted effort has been made over the last decade to grow isotopically enriched silicon crystals, with \SI{99.9}{\percent} and higher content of only one isotope\cite{Bulanov2000,Itoh2003,Ager2005}, for use in a variety of fields such as metrology\cite{Becker2003} and quantum computing\cite{Kane1998}. The availability of such isotopically enriched silicon is of great interest for spectroscopy since it can significantly reduce inhomogeneous spectral linewidths and therefore improve spectral resolution. For example, two recent reports have resolved new fine structures in the optical spectra of phosphorus donors\cite{Cardona2005} and in the electron spin resonance (ESR) spectra of boron acceptors\cite{Tezuka2010} in $^{28}$Si. These fine spectral structures were unresolvable in natural silicon because of the inhomogeneous broadening arising from the presence of magnetic $^{29}$Si nuclei. Furthermore, the absence of $^{29}$Si nuclei in isotopically enriched $^{28}$Si silicon, eliminates the spectral diffusion decoherence mechanism for donors in silicon\cite{Tyryshkin2003}, an otherwise dominant source of decoherence in solid-state spin-based quantum computing architectures\cite{Morton2011,Zwanenburg2013}. In this paper, we perform ESR of exchange coupled phosphorus dimers in isotopically purified $^{28}$Si. The improved spectral resolution arising from isotopic enrichment enables us to observe a previously unresolved fine structure in the dimer ESR line. We then perform pulsed ESR experiments and place limits on the decoherence of phosphorus dimers arising from the presence of exchange coupling.

\begin{figure}
\includegraphics{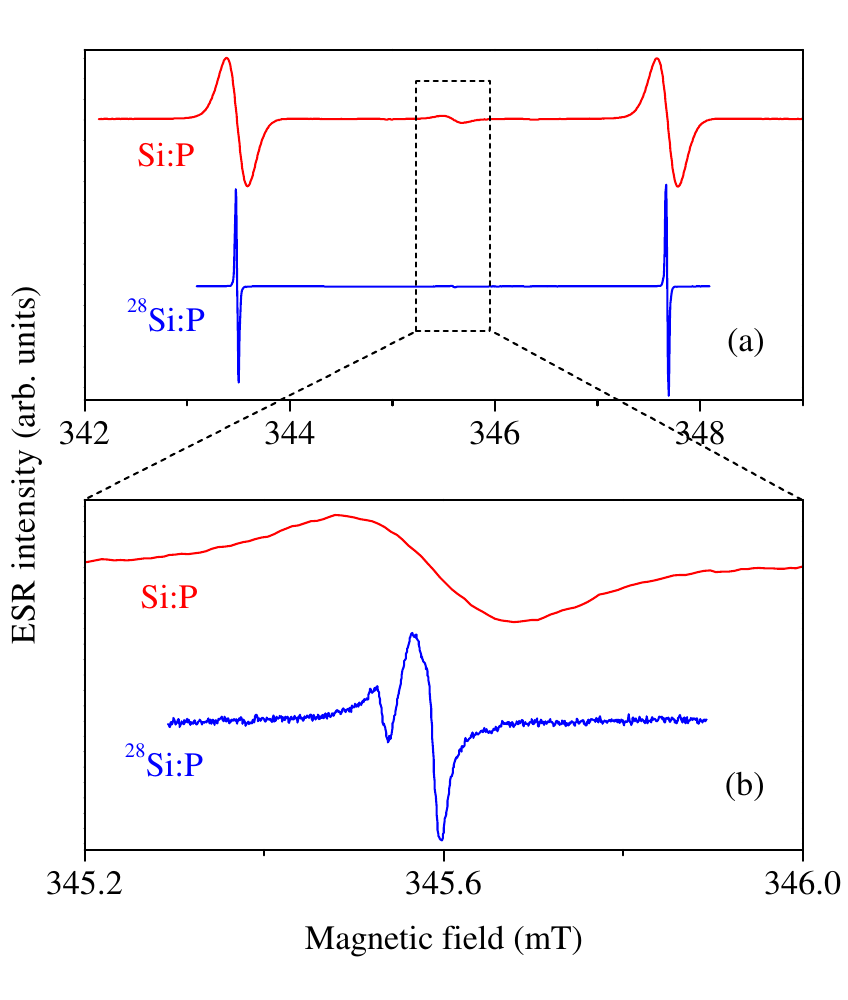}
\caption{(a) ESR spectra from phosphorus dopants in natural silicon (Si:P) and isotopically enriched $^{28}$Si ($^{28}$Si:P), measured at \SI{15}{\kelvin}. Phosphorus doping densities were $\sim$\SI{2E16}{\per\cmc} (see other details of the samples in the text). Two strong lines split by \SI{4.2}{\milli\tesla} are from isolated donors. Weak line in the center is from donor dimers. Two outer lines from the dimers overlap with the stronger lines from isolated donors and are therefore unobservable. (b) Zoom in to the center region showing details of the central line from dimers. \label{fig:spectra}}
\end{figure}

Dopants in natural silicon have been studied for over fifty years by ESR spectroscopy\cite{Feher1959,*FeherGere1959,*Wilson1961}. A typical continuous wave (CW) ESR spectrum of phosphorus donors in natural silicon is shown in Fig.~\ref{fig:spectra}(a). The spectral lines are inhomogeneously broadened by hyperfine interactions with the \SI{4.7}{\percent} of $^{29}$Si (nuclear spin $I=1/2$)\cite{Feher1959,Abe2010}. On the other hand, the spectrum measured in isotopically enriched $^{28}$Si, also shown in Fig.~\ref{fig:spectra}(a), demonstrates a reduced spectral linewidth (the $^{28}$Si nucleus has no magnetic moment, $I=0$). In moderately doped samples ($\sim \SI{e16} {\per\cmc}$), a weak line is also observed at the center of the donor doublet that arises from pairs of donors that are close enough to form exchange ($J$) coupled dimers\cite{Feher1955,Slichter1955}. Since the crystals are randomly doped, the dimers are present with a broad range of distances between dopants and therefore a broad distribution of $J$ couplings. The central ESR line arises from a subset of dimers with $J$ greater than the hyperfine coupling ($A = \SI{117}{MHz} \equiv \SI{4.2}{\milli\tesla}$) to the donor $^{31}P$ nucleus\cite{Cullis1970}. In Fig.~\ref{fig:spectra}(b), we show that the reduced linewidth in $^{28}$Si allows us to observe a new fine structure in the dimer ESR line with a splitting of about \SI{60}{\micro\tesla}. Since $J$ is greater than $A$, the eigenstates of the dimer consist of a spin-0 singlet ($S$) and three spin-1 triplet states ($T_+$, $T_0$, $T_-$) and our ESR experiment probes transitions among the three triplet states. Below, we will explain through simulations that this new fine structure is a result of the second order mixing between the $S$ and $T_0$ states. This mixing arises from an interplay between the $^{31}$P hyperfine, exchange and Zeeman energies, with the mixing being strongest when the exchange coupling is approximately equal to the Zeeman energy. Further, we will show that this mixing allows the use of ESR transitions between the triplet states to probe spin relaxation between the $S$ and $T_0$ triplet state, even though the $S$ state is ESR-silent in our experiments.

We performed pulsed ESR to measure spin relaxation times $T_1$ and $T_2$ of dimers, in order to examine whether the presence of exchange coupling within the dimer gives rise to decoherence in excess of that for the isolated donor. The exchange interaction between two phosphorus donors is the canonical method to implement a two-qubit gate \cite{Kane1998,Vrijen2000,Zwanenburg2013}, and therefore any additional decoherence arising from exchange would be of critical importance for donor-based quantum computing schemes. Such decoherence can be caused by charge noise modulating the $J$-coupling, which would be an especially significant issue for donors near an interface and next to the control gates\cite{Culcer2009}. Furthermore, the $S$--$T_0$ state mixing arising from $J$-coupling opens an additional relaxation ($T_1$) pathway through the electron-phonon interaction\cite{XuedongHu2010}. These mechanisms could thus limit the usability of exchange to perform multi-qubit operations. In our experiments, the ability to resolve a fine structure in the dimer ESR line enables us to separately address two dimer sub-ensembles, the first with $J$ ranging from $6$ to $60$~GHz and the second with $J$ ranging from $2$ to $7$~GHz. We find that the relaxation times of these two dimer sub-ensembles are identical to that of isolated donors, thus limited by the same mechanism, namely spectral diffusion due to dipolar interactions with flip-flopping neighboring donors\cite{Tyryshkin2011}. Thus, we find that the presence of $J$-coupling of $2$--\SI{60}{\GHz} in dimers does not introduce any additional decoherence in bulk Si on a timescale of \SI{10}{\ms}.

An ESR experiment directly probes the spin-dynamics among the triplet ($T$) states of the dimer. However, high-fidelity two-qubit gates require maintaining the coherence between all four spin states including the $S$ state. Although our ESR experiments do not allow any direct information about the ESR-silent $S$ state, nevertheless some important estimates can be made from $T_2$ times measured for $T$ states. Specifically, the one-phonon $T_1$ process from $T_0$ to $S$, arising from the $J$-induced mixing\cite{XuedongHu2010}, can result in an irreversible leakage of coherence from the triplet state during Hahn echo experiments. This would potentially result in faster $T_2$ decay times, which we do not see in our experiment. Our measured $T_2 = 4$~ms for triplet states thus implies that $T_{S-T0}$ must be slower than \SI{10}{\ms}. This lower bound, consistent with recent theoretical calculations\cite{XuedongHu2010} and experimental results\cite{Dehollain2014}, is encouraging for spin-based quantum computing schemes since it suggests that the presence of exchange coupling does not cause any additional relaxation between the triplet and singlet states at the level of about \SI{10}{\ms} in a bulk crystal.

Finally, while measuring $T_2$ using a standard Hahn echo experiment\cite{Hahn1950}, we observed an unusual dependence of the echo decay on pulse lengths and powers. Through numerical simulations we explain that this dependence arises from electron spin echo envelope modulation (ESEEM) effects in high-spin dimers\cite{Morton2005} and destructive interference of the ESEEM from different dimers  due to a broad $J$-coupling distribution in the ensemble. These ESEEM effects, while creating additional difficulties in our experiments, can be viewed as a spectroscopic signature of $J$-coupled dimers and as such can be useful in studying donor dimers and also other coupled dimers.

\section{Experiment}
\label{sec:Experiment}

ESR experiments were performed using isotopically enriched $^{28}$Si epi-layer wafers (\SI{99.9}{\percent} of $^{28}$Si and \SI{800}{ppm} of $^{29}$Si) doped with phosphorus to a density of \SI{1.6E16}{\per\cmc}\cite{Ager2005}. For comparison, we also measured a natural silicon crystal doped with phosphorus to a density of \SI{1.7E16}{\per\cmc}. These samples were previously used in our pulsed ESR study of isolated donors at temperatures down to \SI{7}{\kelvin} in Ref.~\onlinecite{Tyryshkin2003}, where these samples were labeled as ``$^{28}$Si:P-\num{e16}'' and ``Si:P-\num{e16}'', respectively. The CW ESR spectra shown in Fig.~\ref{fig:spectra} were measured using these silicon samples.

\begin{figure}
\includegraphics{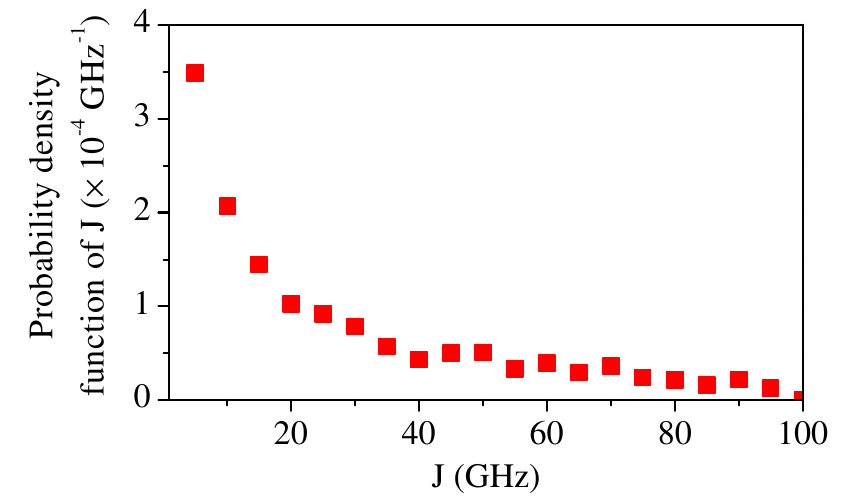}
\caption{Distribution function of dimer $J$ couplings calculated for a doping density \SI{2E16}{\per\cmc}. Only high $J$ range between \SI{1}{\GHz} and \SI{100}{\GHz} is shown specific to dimers selectively observed in our experiments. The complete distribution function, including the low $J$ range, is shown in Fig.~\ref{fig:logJdistribution} in the Appendix.
\label{fig:Jdistribution}}
\end{figure}

The random distribution of dopants in silicon implies that dimers are present with a broad range of $J$ couplings\cite{Cullis1970}. We modified previously used methods to calculate the distribution of $J$ for our dopant density of about \SI{2E16}{\per\cmc}. Details of this calculation are presented in the Appendix where we compare our result with the previous results in Ref.~\onlinecite{Cullis1970}. Only a small fraction of the donor pairs that have $J$ greater than the $^{31}$P hyperfine coupling of \SI{117}{\MHz} contribute to the central dimer line in the ESR experiment\cite{Slichter1955}. In Fig.~\ref{fig:Jdistribution}, we show the calculated probability density function for $J$ of those dimers that we selectively observe in our experiments. Our pulsed ESR experiments are sensitive to an even narrower range of $J$. As discussed below, a majority of our pulsed experiments were performed on the high-field component of the dimer fine structure, centered at \SI{345.63}{\milli\tesla}. In this case about $90$\% of the measured signal arises from a sub-ensemble of dimers with $J$ ranging from about $6$ to $60$~GHz. We also performed an experiment \SI{60}{\micro\tesla} below, on the low-field component of the fine structure, for which most of the signal arises from a dimer sub-ensemble with $J$ ranging from about $2$ to $7$~GHz.

CW and pulsed ESR measurements were performed at X-band (\SI{9.66}{\GHz}) with a Bruker Elexsys 580 ESR spectrometer in a Bruker MD-5 dielectric resonator. An Oxford CF935 helium flow cryostat was used to maintain temperatures down to \SI{5}{\kelvin}, while a temperature of \SI{1.7}{\kelvin} was achieved by filling the cryostat with liquid helium and pumping. As $T_1$ was found to be exponentially dependent on temperature, the sample temperature was precisely controlled to within \SI{0.1}{\kelvin} using an Oxford ITC503 temperature controller.

$T_1$ and $T_2$ were measured using the standard inversion recovery and Hahn echo pulse sequences\cite{Schweiger2001}. In the inversion recovery sequence ($\pi$ -- $t$ -- $\pi$/2 -- $\tau$ -- $\pi$ -- $\tau$ -- echo), the echo intensity was measured as a function of delay, $t$, after the initial inverting $\pi$ pulse. The measured intensity was fit by an exponential dependence to give the characteristic time $T_1$. Similarly, in the 2-pulse Hahn echo sequence ($\pi$/2 -- $\tau$ -- $\pi$ -- $\tau$ -- echo), the echo intensity measured as a function of total time 2$\tau$ was fit with an exponential decay to give the characteristic time $T_2$. In all pulsed experiments, the echo signal intensity was integrated using a \SI{800}{\ns} window symmetrically positioned on top of the echo signal. This integration corresponds to applying a detection bandwidth of \SI{1.2}{\MHz} (\SI{40}{\micro\tesla}), i.e.\ only those dimers which have their resonance field within $\pm$\SI{20}{\micro\tesla} of the applied magnetic field were detected. Finally, a 16-step phase cycling sequence was used to remove any extraneous signals arising from microwave pulse imperfections (e.g., free induction decay signals) that could contaminate the echo decays. In order to eliminate contributions from broad background signals, the echo decays were also measured at a field \SI{0.5}{\milli\tesla} higher than the dimer line and the resulting background decay was subtracted from the dimer decays before fitting to extract the relaxation times.

In order to compare the dimer T$_2$ with those of isolated donors, we also measured T$_2$ for isolated donors in the $^{28}$Si sample at temperatures down to \SI{1,7}{\kelvin}, thus extending the results of Ref.~\onlinecite{Tyryshkin2003} to lower temperature. As in Ref.~\onlinecite{Tyryshkin2003}, the effect of instantaneous diffusion\cite{Klauder1962,*Mims1968} on the $T_2$ relaxation of isolated donors was removed by performing a series of 2-pulse experiments ($\pi$/2 -- $\tau$ -- $\theta_2$ -- $\tau$ -- echo) with reduced rotation angle ($\theta_2$) for the second pulse, and extrapolating the measured $T_2$'s to that which would be measured if $\theta_2$ was zero. This extrapolated $T_2$ then measures the $T_2$ for isolated donors without instantaneous diffusion.

\section{Results}

In the following sections we describe our experimental results and provide their interpretation. The new fine structure resolved in the dimer ESR signal in $^{28}$Si is described in Sec.~\ref{sec:CW expt sim}. The $T_1$ and $T_2$ results are discussed in Sec.~\ref{sec:T1, T2}. While measuring the Hahn echo from dimers, we found an unusual dependence of the echo decay on the pulse lengths and powers. This behavior is explained through ESEEM simulations in Sec.~\ref{sec:ESEEM expt sim}.

\subsection{Fine structure in the dimer ESR lineshape}
\label{sec:CW expt sim}
The full spin Hamiltonian for a dimer in a magnetic field, $B_0$, directed along the $z$-axis, expressed in frequency units\cite{Cullis1970} is 
\begin{eqnarray}
\mathcal{H} = & \nu_e(S_{1z}+S_{2z}) + J(\bm{S_1}\cdot\bm{S_2}) +                 
                                             A(\bm{S_1}\cdot\bm{I_1}+\bm{S_2}\cdot\bm{I_2}) + \nonumber \\
                    & +\nu_n(I_{1z}+I_{2z}).
\label{eq:Ham}
\end{eqnarray}
In this equation, $\bm{S_i}$ and $\bm{I_i}$ are the electron and nuclear spins of two donors ($i=1,2$) forming a dimer ($I = S = 1/2$ for phosphorus donors); $\nu_e = g\mu_B B_0$ is the electron Larmor frequency ($\sim \SI{9.66}{\GHz}$), with $g=1.9985$ being the electron g-factor, and $\mu_B$ the Bohr magneton; $J$ is the exchange coupling between donors; $A$ is the $^{31}$P hyperfine coupling ($\sim \SI{117}{\MHz}$); and $\nu_n$ is the nuclear Larmor frequency ($\sim \SI{6}{\MHz}$). 

In the product basis $\mid S_{1z}S_{2z}I_{1z}I_{2z}\rangle$, the hyperfine coupling term can be split into a diagonal part $A(S_{1z}I_{1z}+S_{2z}I_{2z})$ and an off-diagonal part
\begin{equation}
\mathcal{H}_{off} = A/2(S_{1+}I_{1-}+S_{1-}I_{1+}+S_{2+}I_{2-}+S_{2-}I_{2+}).
\label{eq:off-diag}
\end{equation}
Assuming $|\nu_e-J| \gg A$ (i.e. $J$ far away from $\nu_e$), the off-diagonal part of the hyperfine coupling can be neglected, and $\mathcal{H}$ can be diagonalized analytically\cite{Cullis1970}. The eigenstates and eigenvalues corresponding to this case are displayed in Table~\ref{tab:e-states, e-values}, where uppercase letters ($S$, $T_+$, $T_0$, $T_-$) denote the electron singlet and triplet states, while lowercase letters ($s$, $t_+$, $t_0$, $t_-$) denote the nuclear singlet and triplet states. Including the off-diagonal terms (Eqn.~\ref{eq:off-diag}) results in additional mixing of the eigenstates and second-order shifts to the eigenvalues of order $A^2/|\nu_e\pm J|$ or $A^2/J$\cite{XuedongHu2010}.

\begin{table}
\caption{\label{tab:e-states, e-values} Approximate eigenstates and eigenvalues of the dimer spin Hamiltonian (Eqn.	~\ref{eq:Ham}), ignoring the off-diagonal terms resulting from the hyperfine coupling (Eqn.~\ref{eq:off-diag}).}

\begin{ruledtabular}
\begin{tabular}{lc}
Eigenstates & Eigenvalues \\
\hline \\
$\mid T_+,t_+\rangle$ & $\nu_e + J/4 + A/2 + \nu_n$	\\
$\mid T_+,t_0\rangle,\mid T_+,s\rangle$ & $\nu_e + J/4$ \\
$\mid T_+,t_-\rangle$ & $\nu_e + J/4 - A/2 - \nu_n$	\\
\hline \\
$\mid T_0,t_+\rangle$ & $J/4 + \nu_n$ \\
$\mid T_0,t_0\rangle,\mid T_0,s\rangle$ & $-J/4 +1/2\sqrt{J^2+A^2}$\\
$\mid T_0,t_-\rangle$ & $J/4 - \nu_n$ \\
\hline \\
$\mid T_-,t_+\rangle$ & $-\nu_e + J/4 - A/2 + \nu_n$	\\
$\mid T_-,t_0\rangle,\mid T_-,s\rangle$ & $-\nu_e + J/4$ \\
$\mid T_-,t_-\rangle$ & $-\nu_e + J/4 + A/2 - \nu_n$ \\
\hline \\
$\mid S,t_+\rangle$ & $-3J/4 + \nu_n$ \\
$\mid S,t_0\rangle,\mid S,s\rangle$ & $-J/4 -1/2\sqrt{J^2+A^2}$\\
$\mid S,t_-\rangle$ & $-3J/4 - \nu_n$ \\
\end{tabular}
\end{ruledtabular}
\end{table}

In an ESR experiment, the only transitions excited are those that flip the electron state ($\Delta M_S = \pm 1$) while preserving the nuclear state ($\Delta M_I = 0$). Therefore, eight ESR transitions are allowed between the twelve levels in Table~\ref{tab:e-states, e-values}. These allowed transitions can be further combined into four pairs of transitions which have identical transition frequencies. Two pairs, $\mid T_\pm,t_+ \rangle \leftrightarrow \mid T_0,t_+\rangle$ and $\mid T_\pm,t_- \rangle \leftrightarrow \mid T_0,t_-\rangle$, with the nuclear spins in $\mid t_+\rangle$ and $\mid t_-\rangle$ states, respectively, have their transition frequencies coinciding with the transition frequencies of isolated donors; these transitions are unobservable in our experiments because they overlap with the much stronger transitions from isolated donors (e.g.\ the two intense lines in Fig.~\ref{fig:spectra}(a)). On the other hand, two other pairs, $\mid T_\pm,s\rangle \leftrightarrow \mid T_0,s\rangle$ and $\mid T_\pm,t_0 \rangle \leftrightarrow \mid T_0,t_0\rangle$, with the nuclei in $\mid s \rangle$ and $\mid t_0\rangle$ states, respectively, have distinct transition energies and contribute to the central line in the dimer ESR spectrum in Fig.~\ref{fig:spectra}(b).

\begin{figure*}
\includegraphics{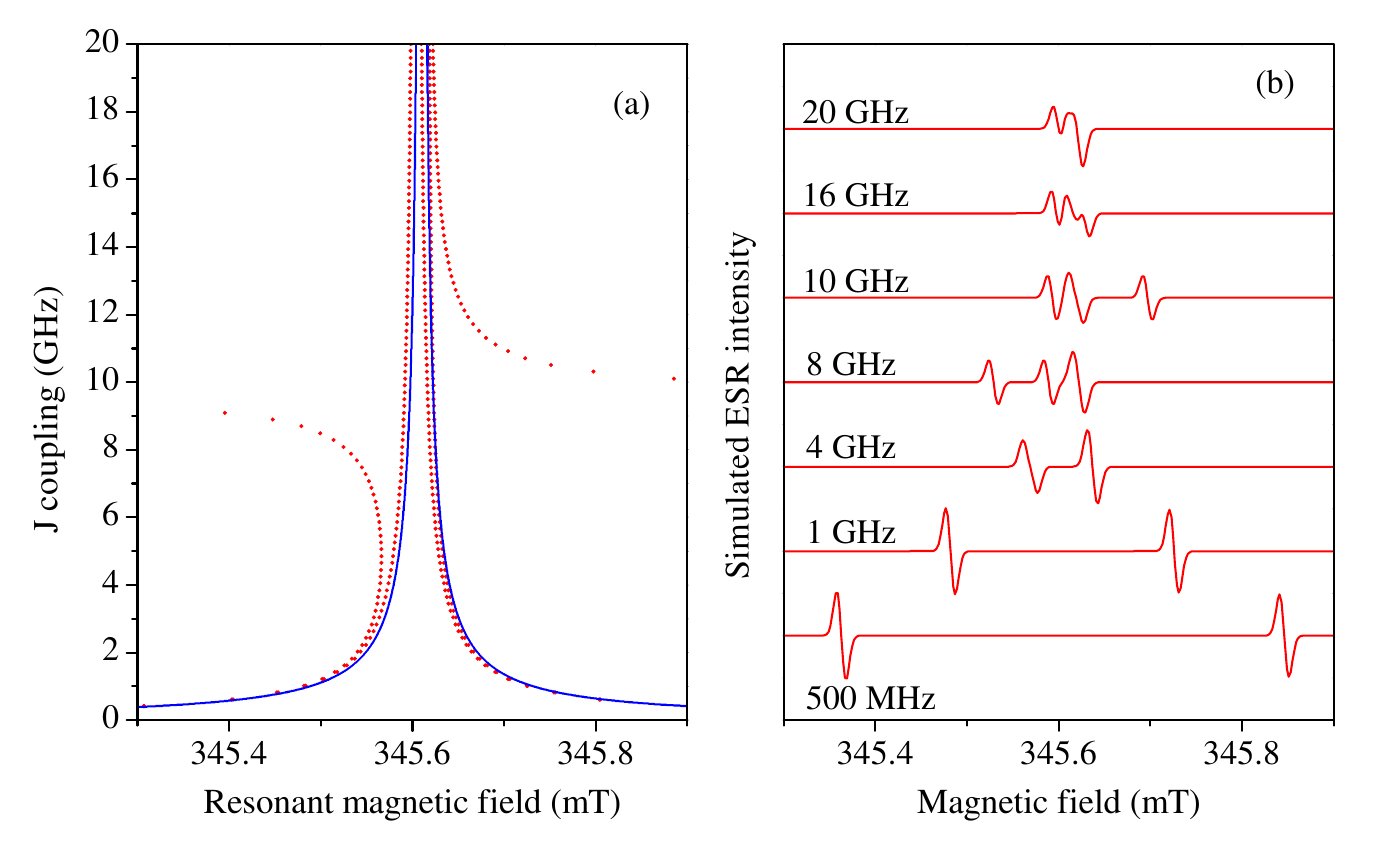}
\caption{(a) Resonant magnetic fields for four central ESR transitions in dimers calculated as function of $J$ coupling, assuming the microwave ESR frequency of \SI{9.66}{\GHz} as in experiment. Note that $J$ couplings are shown as a vertical axis and the calculated resonant fields as a horizontal axis. (Solid blue lines) were calculated using the analytical approximation, ignoring the non-diagonal hyperfine terms (Eqn.~\ref{eq:off-diag}) in the spin Hamiltonian. (Dotted red lines) were calculated numerically, by solving the full spin Hamiltonian $\mathcal{H}$ including the off-diagonal terms. (b) ESR spectra simulated in \textit{EasySpin}\cite{Stoll2006} for dimers with different $J$ couplings (indicated on each spectrum). These simulations assume a microwave frequency of \SI{9.66}{\GHz}, an inhomogeneous spectral linewidth of \SI{10}{\micro\tesla} and a temperature of \SI{15}{\kelvin}. Other simulation parameters are as defined in Eqn.~\ref{eq:Ham}.
\label{fig:transition energies}}
\end{figure*}

For a given $J$, the corresponding transition frequencies for the $\mid s \rangle$ and $\mid t_0\rangle$  transition pairs are $\nu_e \pm (\sqrt{J^2 + A^2}/2 - J/2)$\cite{Cullis1970}, and thus they are symmetrically positioned around the center $\nu_e$, split by $\sim A^2/2J$. These transitions are plotted as resonant magnetic fields in Fig.~\ref{fig:transition energies}(a) (solid blue lines) where for easier comparison with the experimental spectra we show $J$ couplings on a vertical axis and the resonant magnetic fields calculated for each $J$ on a horizontal axis. The symmetric positioning of the resonant fields for all $J$'s implies that after summing over the $J$ distribution as in our samples, the resulting ESR line will be symmetric in contrast to the asymmetric line observed in the experiment (Fig.~\ref{fig:spectra}). Therefore, the asymmetric structure must be an effect of the neglected off-diagonal components in the spin-Hamiltonian (Eqn.~\ref{eq:off-diag}) that mix the singlet and triplet electron and nuclear eigenstates for those dimers having $J \sim \nu_e$. 

To verify this effect, we numerically solved the full spin Hamiltonian $\mathcal{H}$, including the off-diagonal terms. The dotted lines in Fig.~\ref{fig:transition energies}(a) show the positions of the four transitions near the center calculated for different values of the $J$ coupling. As expected, the analytic approximation that gives a symmetric splitting $\sim A^2/2J$ fails for $J$ near $\nu_e = \SI{9.66}{\GHz}$. Instead, the off-diagonal terms mix the singlet and triplet electron and nuclear eigenstates for dimers, thus introducing the asymmetric splitting in the resonant fields. Specifically, the triplet state $|T_-,s\rangle$ is mixed with the state $|S,t_-\rangle$ when $J$ is close to  $\nu_e$ resulting in a doublet transition that is asymmetric about $\nu_e$. To further illustrate this asymmetric splitting, we show in Fig.~3b the CW ESR spectra simulated using the \textit{EasySpin} package\cite{Stoll2006}. The off-diagonal Hamiltonian terms cause an asymmetric splitting of the transitions about the center, with the largest effect seen when $J$ is close to the Larmor frequency.

\begin{figure}
\includegraphics{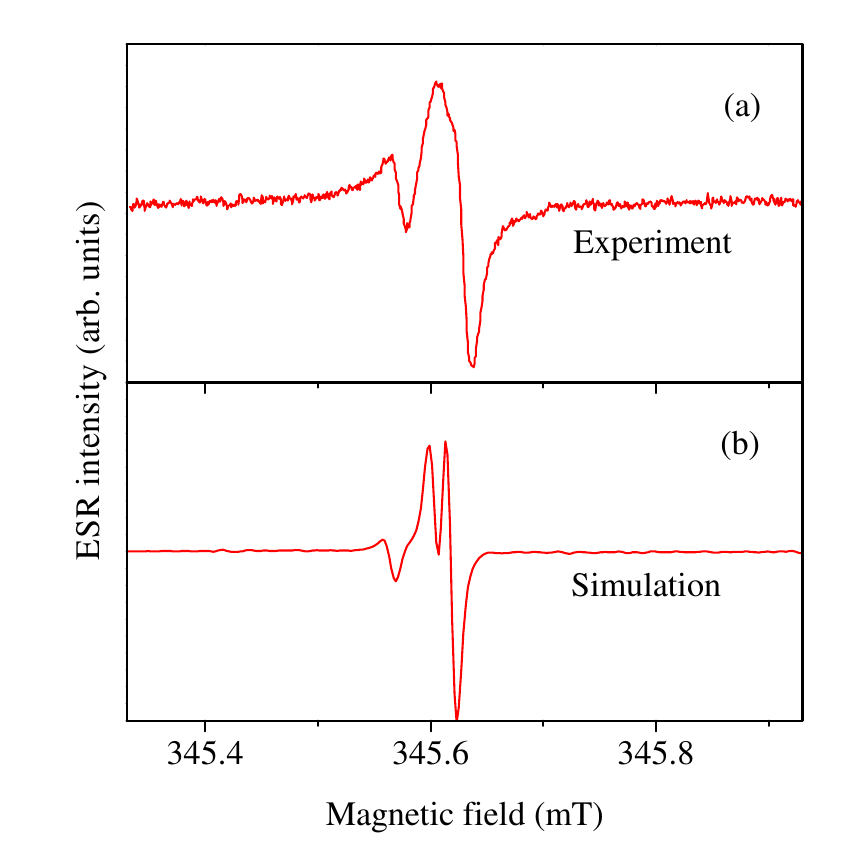}
\caption{Comparison of the experimental (a) and simulated (b) CW ESR signals of phosphorus dimers in doped $^{28}$Si. The experimental spectrum was measured at \SI{15}{\kelvin}. The simulated spectrum was calculated for $N_d = \SI{2E16}{\per\cmc}$, summing the individual spectra over the $J$ distribution function shown in Fig.~\ref{fig:Jdistribution} and assuming $T = \SI{15}{\kelvin}$.
\label{fig:cw comparison}}
\end{figure}

Since our sample contains dimers with a broad range of $J$ values, the final ESR spectrum comprises a sum of the individual spectra for each $J$, weighted by the probability distribution of $J$ as shown in Fig.~\ref{fig:Jdistribution}. The final simulated CW ESR spectrum is shown in Fig.~\ref{fig:cw comparison}), demonstrating excellent agreement with the experiment in the position of the dimer line and its asymmetric structure. The relative intensity of the low-field sub-structure is somewhat lower in the simulation as compared to the experiment. An extra splitting can also be observed on the high-field line in the simulated spectrum. We ascribe these to the inaccuracy of the estimated $J$ distribution function (Fig.~\ref{fig:Jdistribution}). Nevertheless, we can conclude that overall the simulations are in good qualitative agreement, and thus we understand the origin of the fine structure of the phosphorus dimer line in $^{28}$Si.

\subsection{$T_1$ and $T_2$ of dimers, and comparison to isolated donors}
\label{sec:T1, T2}

Most of our pulsed ESR measurements were done with the magnetic field centered on the larger high-field structure in the dimer line (referred to as line\#1, at \SI{345.63}{\milli\tesla} in Fig.~\ref{fig:cw comparison}(a)). The spectral bandwidth in our pulsed experiments was \SI{40}{\micro\tesla}, and therefore referring to Fig.~\ref{fig:transition energies}(a) we infer that dimers with $J > \SI{6}{\GHz}$ contributed to the measured relaxation decays, involving all four central transitions, $\mid T_\pm,s\rangle \leftrightarrow \mid T_0,s\rangle$ and $\mid T_\pm,t_0 \rangle \leftrightarrow \mid T_0,t_0\rangle$, with the dimer $^{31}$P nuclei in $\mid s \rangle$ and $\mid t_0\rangle$ states. Furthermore, as seen from the probability distribution of $J$ couplings in Fig.~2, 90\% of the signal arises from dimers with $J$ less than about 60~GHz. 

\begin{figure}
\includegraphics{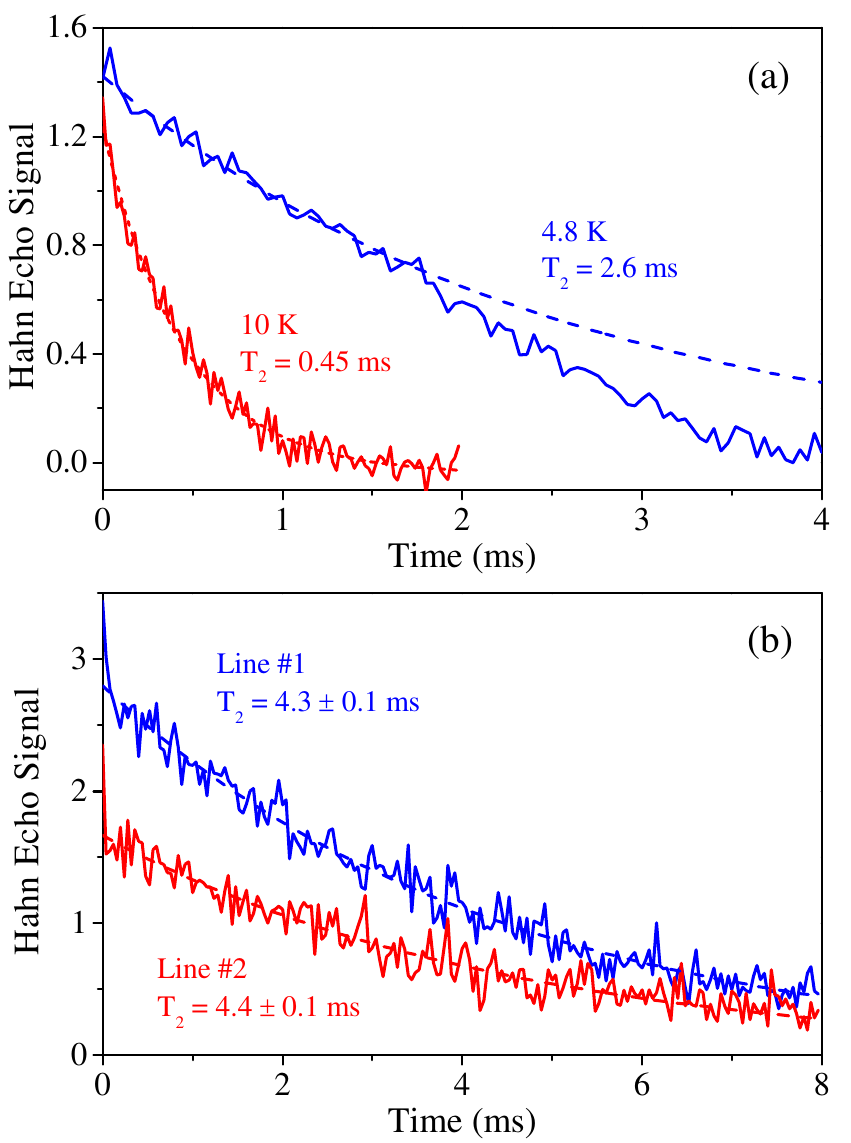}
\caption{(a) Hahn echo decays (solid lines) measured by conventional averaging (see text) on the main dimer line at \SI{345.63}{\milli\tesla} (a sub-ensemble with $ \SI{6}{\GHz} < J < \SI{60}{\GHz}$) in $^{28}$Si at $T = 4.8$ and $\SI{10}{\kelvin}$, with the respective exponential fits (dashed lines) to extract $T_2$. Faster than exponential decay after time of $2$~ms is an artifact of signal averaging in the presence of echo phase fluctuation due to magnetic field noise\cite{Tyryshkin2006b}. (b) Hahn echo decays (solid lines) with exponential fits (dashed lines), measured at $T = 1.7$~K by magnitude detection (see text). Line\#1 (blue) corresponds to the main dimer line at \SI{345.63}{\milli\tesla} and line\#2 (red) corresponds to the satellite line at \SI{60}{\micro\tesla} lower field. The sharp initial drop in the echo signal is an artifact of the destructive interference from partially suppressed ESEEM effects, as discussed further in Sec.~\ref{sec:ESEEM expt sim}
\label{fig:echo decay}}
\end{figure}

Fig.~\ref{fig:echo decay}(a) shows Hahn echo intensity as a function of the total decay time ($2\tau$), at two temperatures. At temperatures between $8$ to \SI{4.8}{\kelvin}, the echo decay was distorted after \SI{2}{\milli\second} decay time into a non-exponential decay, similar to that observed earlier for isolated donors in this sample\cite{Tyryshkin2003}. Such non-exponential decays arise from averaging single-quadrature echo intensities, which fluctuate in phase due to ubiquitous magnetic field noise\cite{Tyryshkin2006b,Biercuk2009,Saeedi2013}. The distorted decay for times greater than \SI{2}{\ms} was ignored, and only the initial part of the decay was used to extract $T_2$ for measurements down to \SI{4.8}{\kelvin}.

At $T=\SI{1.7}{\kelvin}$, the spin polarization is large enough that we can measure an echo signal in a single-shot, i.e.\ without averaging. This enables us to average the magnitude of the echo intensity, rather than a single-quadrature, and thus counters the phase fluctuation due to field noise\cite{Tyryshkin2006b}. The echo decay measured by such a magnitude detection technique, shown in Fig.~\ref{fig:echo decay}(b), displays a simple exponential decay. At \SI{1.7}{\kelvin}, the echo decay was also measured on the low-field satellite line (line \#2 at \SI{345.57}{\milli\tesla}), where as seen from Fig.~\ref{fig:transition energies}, only the transition $\mid T_-,s\rangle \leftrightarrow \mid T_0,s\rangle$ is excited for  dimers with $J$ ranging from $2$ to \SI{7}{\GHz}. We find that both dimer sub-ensembles display the same characteristic $T_2$ of about \SI{4}{\ms} at \SI{1.7}{\kelvin}.

\begin{figure}
\includegraphics{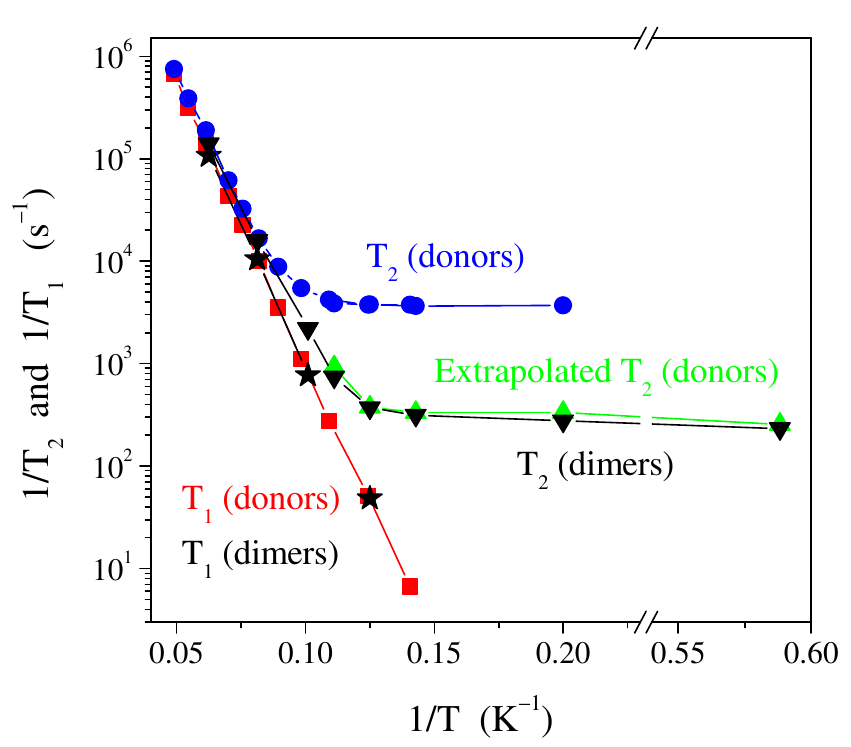}
\caption{Temperature dependence of $T_1$ and $T_2$ for donors and dimers in $^{28}$Si: (stars) dimer $T_1$, (inverted triangles) dimer $T_2$, (squares) donor $T_1$, (circles) donor $T_2$, and (triangles) donor $T_2$ after suppressing instantaneous diffusion. The donor $T_1$ and $T_2$ data down to \SI{7}{\kelvin} are reproduced from Ref.~\onlinecite{Tyryshkin2003}. The lines are guides for eye.
\label{fig:T1, T2 vs T}}
\end{figure}

The temperature dependence of $T_1$ (stars) and $T_2$ (inverted triangles) for dimers is summarized in Fig.~\ref{fig:T1, T2 vs T} along with the corresponding data for isolated donors (squares and circles), measured in the same $^{28}$Si sample. The dimer $T_1$ is identical to that of donors down to $8$~K and is therefore controlled by the same Orbach relaxation mechanism\cite{Castner1967} in this temperature range. While we did not measure the dimer $T_1$ at lower temperatures, we expect it to follow the isolated donor $T_1$, with $T_1 \sim 1$~hour at $1.25$~K\cite{FeherGere1959}. The dimer $T_2$ follows the $T_1$ dependence at high temperatures but then saturates at the level of about \SI{4}{\ms} below \SI{8}{\kelvin}. On the other hand, the donor $T_2$ measured using a standard Hahn echo experiment saturates at an order of magnitude shorter $T_2 =\SI{0.3}{\ms}$. The difference is explained by instantaneous diffusion\cite{Tyryshkin2003} that limits the standard Hahn echo $T_2$ for isolated donors to \SI{0.3}{\ms} in a sample with density of \SI{1.6E16}{\per\cmc}. The effect of instantaneous diffusion was removed as discussed in Sec.~\ref{sec:Experiment}, and the extrapolated $T_2$ of isolated donors is plotted in Fig.~\ref{fig:T1, T2 vs T} as upright triangles. The extrapolated donor $T_2$ matches the dimer $T_2$, saturating at \SI{4}{\ms}. We have recently shown that the donor $T_2$ at low temperatures (below $8$~K) is limited by spectral diffusion due to the dipolar interaction with flip-flopping neighboring donors\cite{Tyryshkin2011}; it follows from Fig.~\ref{fig:T1, T2 vs T} that the dimer $T_2$ is limited by the same process. 

We note that in the Ref.~\onlinecite{Tyryshkin2011}, we showed that the effect of the spectral diffusion process is reduced in samples with lower dopant density, such that the isolated donor $T_2$ becomes $\sim 1$~s at $10^{14}$~cm$^{-3}$, significantly longer than measured here, though still shorter than the ultimate limit of $2T_1$. It would thus be interesting to measure if the dimer $T_2$ can also reach $1$~s timescales. Such a measurement by an ensemble ESR experiment would however require a differently prepared sample and an improved sensitivity than our current setup.

To conclude, we find no evidence of $J$ coupling causing additional $T_1$ and $T_2$ processes between the triplet states, $\mid T_\pm,s\rangle \leftrightarrow \mid T_0,s\rangle$ and $\mid T_\pm,t_0 \rangle \leftrightarrow \mid T_0,t_0\rangle$, in dimers with $\SI{6}{\GHz} < J < \SI{60}{\GHz}$ at temperatures 1.7--\SI{20}{\kelvin}. We also found that $J$-coupling does not limit $T_2$ for the $\mid T_-,s\rangle \leftrightarrow \mid T_0,s\rangle$ transition in dimers with $\SI{2}{\GHz} < J < \SI{7}{\GHz}$ between $1.7$--\SI{4.8}{\kelvin}. By measuring the dimer $T_2 $ of $\SI{4}{\ms}$ at \SI{1.7}{\kelvin} and confirming that it is entirely limited by donor flip-flops and thus shows no additional contribution from $J$-coupling related processes, we can then estimate that $T_2$ of coherences between the triplet states of isolated dimers should be in excess of $\SI{10}{\ms}$ at low temperatures.

Pulsed ESR experiments probe only transitions between triplet ($T_0$, $T_\pm$) states, and the singlet ($S$) state remains silent. At first glance, our $T_1$ and $T_2$ data as measured for the triplet states do not provide any direct information about dynamics involving the singlet state. However the mixing between the $S$ and $T_0$ state ($\sim A^2/J$, as shown above) provides a channel for additional decoherence to be observed by ESR\cite{XuedongHu2010}. $T_{(S-T0)}$ processes cause irreversible leakage of spin population from triplet states to the singlet state; this leakage (if fast on the time scale of our ESR experiment) will cause a loss of coherence between the triplet states and thus should directly limit the measured $T_2$. In experiment, we find that $T_2 = \SI{4}{\ms}$ at 1.7~K is limited by donor flip-flop processes and thus the $T_{(S-T0)}$ processes contribute insignificantly on this time scale. Therefore, we estimate a lower bound  for $T_{(S-T0)}$ to be longer than $\SI{10}{\ms}$ in our sample, for all dimers with $\SI{2}{\GHz} < J < \SI{60}{GHz}$ at \SI{1.7}{\kelvin}.  Our $T_{(S-T0)}$ correlates approximately with $T_{(S-T0)} =1\text{--}\SI{100}{\ms}$ as can be derived from theoretical calculations for the one-phonon (direct) relaxation process in donor dimers with $J =2\text{--}\SI{60}{\GHz}$ at $1.7$~K and magnetic field $0.35$~T as in our experiments\cite{XuedongHu2010}.

\subsection{ESEEM effects in dimer's Hahn echo decays}
\label{sec:ESEEM expt sim}

\begin{figure}
\includegraphics{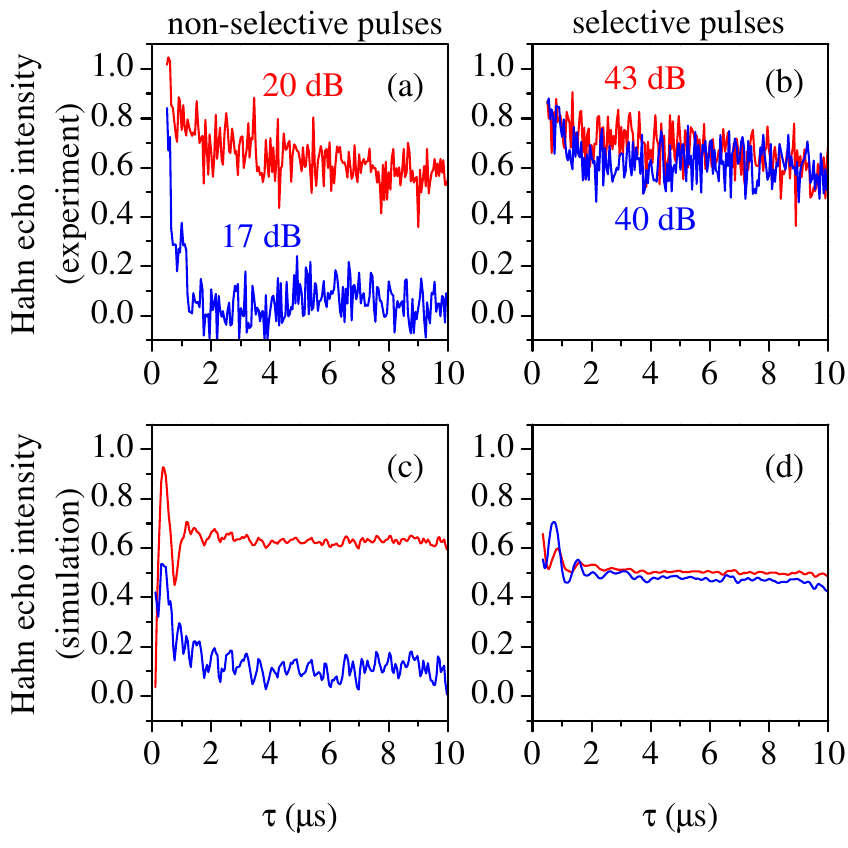}
\caption{Hahn echo decays for dimers at two different settings of pulse durations and pulse powers: (a and b) Experimental decays measured using non-selective (\num{16} and \SI{32}{\nano\second}) and selective (\num{256} and \SI{512}{\nano\second}) microwave pulses, respectively. Two pulse powers (indicated on each trace as an attenuation level from the maximum output power, 1~kW, of the TWT amplifier) were used in each case. With \num{16}-\SI{32}{\nano\second} (\num{256}-\SI{512}{\nano\second}) pulses, the microwave power attenuation level \SI{17}{dB} (\SI{40}{dB}) was selected so as to produce $\pi/2 - \pi$ rotations for isolated donor spins $S = 1/2$.  Measurements were done at \SI{9}{\kelvin}. (c and d) Corresponding simulations using the ESEEM model described in the text.\label{fig:eseem}}
\end{figure}

While measuring dimer's $T_2$, we observed an unusual dependence of the Hahn echo decays on rotation angles of microwave pulses. Fig.~\ref{fig:eseem}(a) illustrates the effect by showing a dramatic change in the echo decay time when using non-selective (broadband) pulses and changing microwave power of the pulses by only 3~dB (corresponds to $\sim 40$\% change in rotation angles of the pulses).
This sharp dependence of the echo decay on a slight change in pulse power is rather unusual and has never been previously reported. Next, we show that this sharp dependence is a result of electron spin echo envelope modulation (ESEEM) arising from the structure of the dimer line. 

ESEEM effects in high-spin electron systems ($S > 1/2$) may arise when non-selective microwave pulses excite multiple electron spin transitions\cite{Yudanov1969,Milov1986,Morton2005}. When these transitions have slightly different resonant frequencies, the spin-populations associated with each transition are interconverted by the non-selective refocusing pulse. This gives rise to a modulation of the echo signal decay, with a modulation frequency determined by the difference between the two excited transition frequencies\cite{Morton2005}. 

As discussed above (Sec.~\ref{sec:CW expt sim}), several transitions, $\mid T_\pm,s\rangle \leftrightarrow \mid T_0,s\rangle$ and $\mid T_\pm,t_0 \rangle \leftrightarrow \mid T_0,t_0\rangle$, from each dimer can contribute to the central dimer line. These transitions are split by $\sim A^2/2J$ (if $|\nu_e-J| \gg A$) which is less than $\SI{40}{\micro\tesla}$ for $J>\SI{6}{\GHz}$. This splitting is smaller than the excitation bandwidth of our non-selective pulses, \SI{0.5}{\milli\tesla} for pulse lengths of 16 and \SI{32}{\ns} used in experiments shown in Fig.~\ref{fig:eseem}(a)). Therefore, all four transitions should be excited by our non-selective pulses, and the resulting echo decay should oscillate with their frequency difference $A^2/2J$. Dimers in our sample are present with a broad range of $J$ and therefore with a broad range of oscillation frequencies. The oscillation frequencies from different $J$ can interfere destructively killing the echo signal at long times. This is exactly what is observed in Fig.~\ref{fig:eseem}(a) when using non-selective pulses (blue trace). The situation changes when reducing the power by 3~dB (red trace). Now, the destructive interference is not as effective, leaving the echo signal to decay for long times.

To verify our ESEEM-related model we repeated Hahn echo experiments using selective microwave pulses (Fig.~\ref{fig:eseem}(b)). The excitation $B_1$ field is set to \SI{35}{\micro\tesla} (for \num{256}-\SI{512}{\ns} microwave pulses at 40~dB attenuation) that is smaller or comparable to typical dimer splittings. Therefore, only one of the transitions is selectively excited in each dimer (in practice other transitions are still excited but to a lesser extent than the selected transition), and thus the resulting echo decays should show no (or greatly reduced) ESEEM oscillations. The destructive interference should be greatly suppressed, and the echo signal should last for longer times. Indeed, this is what we find in experiment (Fig.~\ref{fig:eseem}(b)); both high and low power traces show comparably long decays and no sharp dependence on microwave power is observed as expected.

Before proceeding to numerical simulations of these ESEEM effects, we need to clarify our choice of microwave powers ($B_1$ fields) in the experiments shown in Figs.~\ref{fig:eseem}(a) and (b). For a spin with $S = 1$ as in our dimers, the choice of optimal power (that at given pulse durations produces $\pi/2 - \pi$ rotations and thus a maximal echo signal) depends on whether pulses are selective or non-selective. In case of non-selective pulses that excite both triplet transitions, $\mid T_+\rangle \leftrightarrow \mid T_0\rangle$ and $\mid T_-\rangle \leftrightarrow \mid T_0\rangle$, the optimal power for the dimer spin ($S=1$) is the same as for a $S = 1/2$ spin (e.g. isolated donors). On the other hand, in case of selective pulses that excite only one of the triplet transitions, the optimal power should be a factor of 2 smaller (3~dB lower) because the Rabi frequency for $S=1$ is a factor of $\sqrt{2}$ faster than for $S=1/2$ (Ref.~\onlinecite{Schweiger2001}, chapter 6). In our setup, the powers \SI{17}{dB} (\SI{40}{dB}) for \num{16}-\SI{32}{\nano\second} (\num{256}-\SI{512}{\nano\second}) pulses were calibrated such as to produce $\pi/2 - \pi$ rotations (and a maximal echo signal) for isolated donor spins $S = 1/2$. Thus, the power required to achieve $\pi/2-\pi$ rotations on a dimer selectively or non-selectively were $40$~dB or $20$~dB, respectively. This calibration procedure allows us to connect the power levels in the experiments with the microwave pulse rotation angle to be used for numerical simulations discussed next.

\begin{figure}
\includegraphics{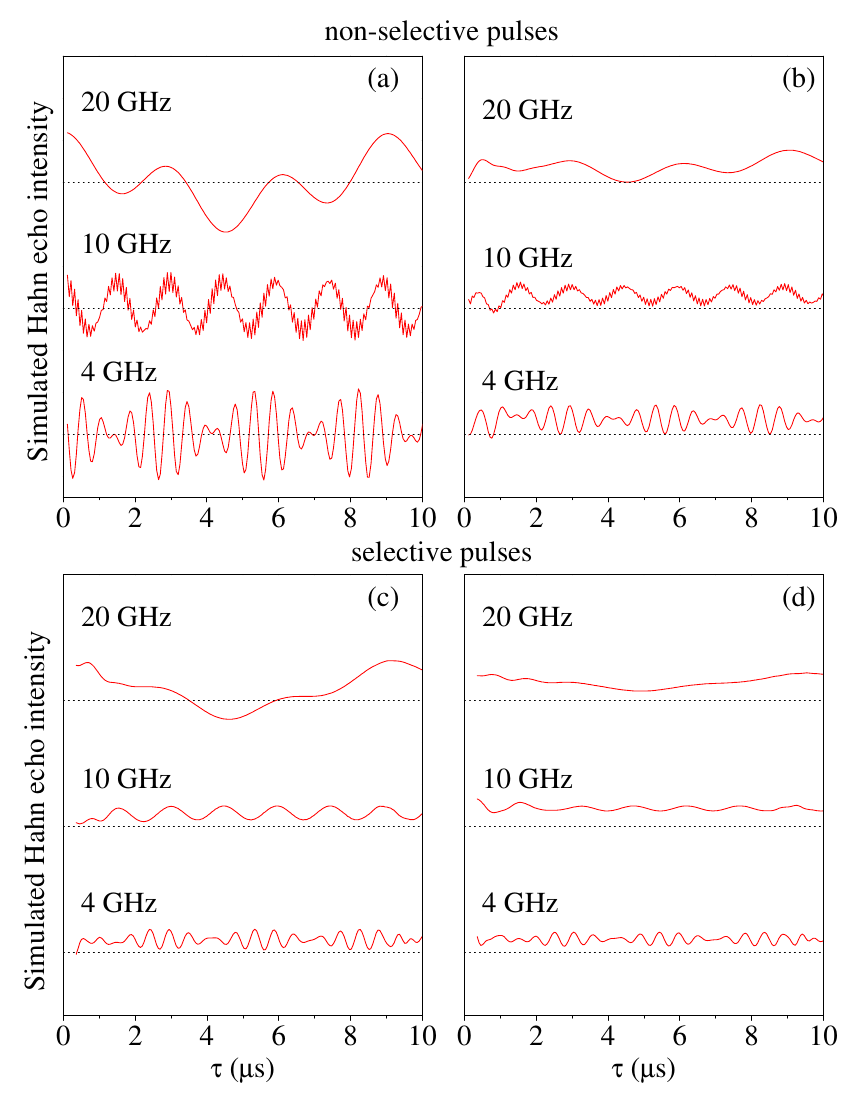}
\caption{Simulated Hahn echo signals for different $J$'s assuming a linewidth of \SI{10}{\micro\tesla}, a magnetic field centered on the high field line of the dimer spectrum and pulse lengths in (a), (b) of \num{16}-\SI{32}{\nano\second} (non-selective pulses) and in (c), (d) of \num{256}-\SI{512}{\nano\second} (selective pulses). The traces are offset for clarity and the zero line for each trace is marked with dashed lines. In (a), pulse power set so that a \SI{16}{\nano\second} pulse rotates a donor spin by $\pi/2$. In (c), pulse power set so that a \SI{256}{\nano\second} pulse rotates a donor spin by $\pi/2$. Pulse power in (b) and (d) are set \SI{3}{dB} lower than (a) and (c) respectively.\label{fig:eseem, diffJ}}
\end{figure}

We numerically simulated the echo signal intensity as a function of $\tau$ for dimers with given $J$, including the magnetic field offset and ESR linewidth as in the experiment. The echo intensity is calculated from the expectation value of $S_y$ traced over the final density matrix $\mathrm{Tr}( U \rho_0 U' S_y)$, where $U$ is the time evolution operator for the pulse sequence and $\rho_0 = S_z$ is the pseudo-pure initial density matrix. $U = U_{f}U_{p2}U_{f}U_{p1}$, where $U_{p1}$ and $U_{p2}$ are the evolution operators during the pulses, and $U_{f}$ is a free evolution operator between the pulses. The free evolution operator is $\exp(-\mathrm{i} 2 \pi \mathcal{H} \tau)$, where $\mathcal{H}$ is the spin-Hamiltonian of a dimer (Eq.~\ref{eq:Ham}), and $\tau$ is the delay between pulses. The pulse operators are $\exp(-\mathrm{i} 2 \pi (\mathcal{H}+\mathcal{H}_{pulse}) t_{p})$, where $\mathcal{H}_{pulse} = \gamma H_1 S_x$ is the Hamiltonian of the microwave pulse, and $t_p$ is the pulse length. Calculations using the full $16\times16$ Hamiltonian of a dimer were found to be extremely demanding of CPU power, especially when taking a large number of time steps and averaging over several parameters. Therefore we sped up the calculation by dividing the full $16\times16$ Hamiltonian matrix  $\mathcal{H}$ into four $3\times3$ matrices (triplet states $S=1$, one for each out of four nuclear spin states) and four $1\times1$ matrices (singlet state $S=0$). We numerically confirmed that the mixing between these matrices is negligible during the spin evolution (less than 0.1\%), and therefore each matrix can be evaluated independently. Only two $3\times3$ matrices need to be evaluated for the 6 dimer eigenstates involved in the central dimer line. The echo signal is calculated by integrating the echo intensity over a window of \SI{800}{\ns} centered on $2 \tau + t_{p1}$, similar to the experiment. The simulated traces were averaged over the inhomogeneous ESR linewidth by including a term $\gamma \Delta H S_z$ in $\mathcal{H}$, where $\Delta H$ is an offset magnetic field as seen by a dimer in the ensemble due to inhomogeneity (a Gaussian linewidth of \SI{10}{\micro\tesla} was assumed as found in CW ESR simulations). 

The representative results of our simulations for $J={4}\text{--}\SI{20}{\GHz}$ are shown in Fig.~\ref{fig:eseem, diffJ}. ESEEM oscillations are observed in all cases, including non-selective and selective pulses and also large and small pulse rotation angles (pulse powers). The oscillation frequency scales proportionally to $A^2/2J$ as expected. The oscillation amplitude is most pronounced when using non-selective pulses in combination with large rotation angles (Fig.~\ref{fig:eseem, diffJ}(a)). The oscillations span the full amplitude range, cross zero and invert the signal to negative. The oscillation amplitude is suppressed by a factor of 2 when using smaller rotation angles (Fig.~\ref{fig:eseem, diffJ}(b)). An important improvement with smaller rotations is that the echo traces contain a non-oscillatory (i.e.\ zero frequency) component. The ESEEM oscillations are further suppressed with selective pulses (Fig.~\ref{fig:eseem, diffJ}(c and d)), as expected because only one of the triplet transitions is selectively excited. Further, the selective pulses at both rotation angles produce a non-oscillatory component in the echo trace.

To calculate the echo signal trace for the dimer ensemble in our silicon crystal we sum up traces like in Fig.~\ref{fig:eseem, diffJ} over the $J$ coupling distribution shown in Fig.~\ref{fig:Jdistribution}. The broad distribution of $J$ couplings implies a broad distribution of ESEEM frequencies. When summed up over this distribution, the oscillating components of the echo signals should average to zero (because of a destructive interference between different frequencies) and only the non-oscillatory components should survive. The summation results are shown in Fig.~\ref{fig:eseem}(c and d). The resulting curves qualitatively correlate with the experimental results. In particular, for non-selective pulses and using large rotation angles, the simulated trace decays quickly to zero within the first two microseconds similar to the experiment at \SI{17}{dB} attenuation. Other traces, including non-selective pulses with small rotation angles and selective pulses at both rotation angles, show a strong signal over a long time range. The residual oscillations seen in all simulated traces are possibly an artifact of averaging over the $J$ distribution (the noise level in the experiment is higher than any of the predicted oscillations). Thus, our simulations confirm that the fast decay of the echo at short $\tau$'s in some experiments is an artifact associated with ESEEM that arises from non-selectively exciting both $\mid T_+\rangle \leftrightarrow \mid T_0 \rangle$ and $\mid T_0\rangle \leftrightarrow \mid T_- \rangle$ transitions of dimers. To remove this artifact and measure the true dimer $T_2$ decays requires using selective pulses.

\section{Conclusion}
\label{sec:Conclusion}
We have resolved a new fine structure in the ESR spectra of exchange coupled phosphorus dimers in $^{28}$Si. This fine structure had not been observed in previous experiments because of the significant inhomogeneous ESR linewidth caused by the presence of $^{29}$Si nuclei in natural silicon samples. We have shown through numerical simulations that this fine structure is a result of singlet-triplet state mixing due to the interplay of the exchange, hyperfine and Zeeman energies, averaged over the broad $J$ distribution of dimers in doped silicon crystals.

We have also measured the $T_1$ and $T_2$ for triplet state transitions of dimers in bulk $^{28}$Si. The $T_2$ experiment is complicated by the presence of ESEEM effects that depend on pulse lengths and powers. By appropriate choice of pulse parameters, we were able to suppress the ESEEM effect and measure the relaxation times. The observed ESEEM effects can be regarded as a spectroscopic signature of $J$-coupled dimers and can be used in other studies of exchange coupled systems. In ensemble experiments, the ESEEM can show up as artificially fast relaxation decays, while in single dimer studies it might reveal observable modulation effects.

We find that spin coherence times among the triplet states in dimers with $\SI{2}{\GHz} < J < \SI{60}{GHz}$ at \SI{1.7}{\kelvin} are not affected by the presence of $J$-coupling at least on time scales up to \SI{10}{\ms}. While we were not able to observe the singlet-to-triplet transitions in dimers directly, we were able to estimate that $T_1$ leakage from the triplet to singlet state in dimers must be longer than \SI{10}{\ms}, consistent with recent theoretical estimates\cite{XuedongHu2010} and experimental results\cite{Dehollain2014}. Note that we are currently unable to put bounds on the decay of coherences between the triplet and singlet state, which is an important topic for further study, since these also limit the fidelity of a two-qubit gate implemented via $J$-coupling. Furthermore, dimers close to an interface also need to be investigated since they may have shortened coherence, as seen for isolated donors\cite{Schenkel2006}.

\appendix*
\section{$J$-coupling distribution for dimers in a silicon crystal with donor concentration of \SI{2E16}{\per\cmc}}
\label{sec: J distribution}
Calculations of the $J$ coupling distributions in phosphorus doped silicon crystals have been previously reported in Ref.~\onlinecite{Cullis1970,Andres1981,New1984,New1985}. We adapted the previously used methods to calculate the $J$ distribution for the donor density used in our experiments. We use a Monte-Carlo technique by repeatedly and randomly positioning dopants in a given volume. For a donor density $N_d$, the side of a cube containing on average one donor is $L = 1/N_d^{1/3}$. We simulate a silicon lattice cube of $5 L\times 5 L \times 5 L$ volume which contains on average 125 donors. Since the donors are distributed uniformly, the number of donors in the cube is a Poisson random variable with mean 125. Assuming one of the donors to be at the origin, in each iteration we generate a random number of dopants from a Poisson process with mean 124. The donors are placed according to a uniform distribution, randomly on the silicon lattice (an FCC lattice with two point basis (000), $a/4$(111); lattice constant $a = \SI{5.43}{\angstrom}$~\cite{AM1976}). 

For each donor configuration, we calculate $J$ couplings between the donor at the origin and all other donors. The expression for exchange coupling, given in the Heitler-London formalism, is taken from Ref.~\onlinecite{Cullis1970}. Similar expressions taking into account the oscillatory dependence of $J$ coupling on donor-to-donor distance were also published in Refs.~\onlinecite{Andres1981,Koiller2001}. The donor whose $J$ coupling is maximum is assumed to form a dimer pair with the donor at the origin. In addition, we take into account the possibility of forming trimers and higher order clusters. Thus, we ignore the dimers where the second donor in the pair is more strongly coupled to a third donor. Further, we ignore dimers in which either donor in the pair is coupled to a third donor with a $J$ greater than \SI{1}{\MHz}. The \SI{1}{\MHz} threshold was chosen since a coupling of this strength would result in a dimer ESR signal splitting greater than the ESR linewidth observed in our experiment. On average, for $N_d$ less than \SI{4E16}{\per\cmc} we found that less than \SI{10}{\percent} of a million iterations were ignored, implying that that about \SI{10}{\percent} of the donors form trimers or higher order clusters. 

\begin{figure}
\includegraphics{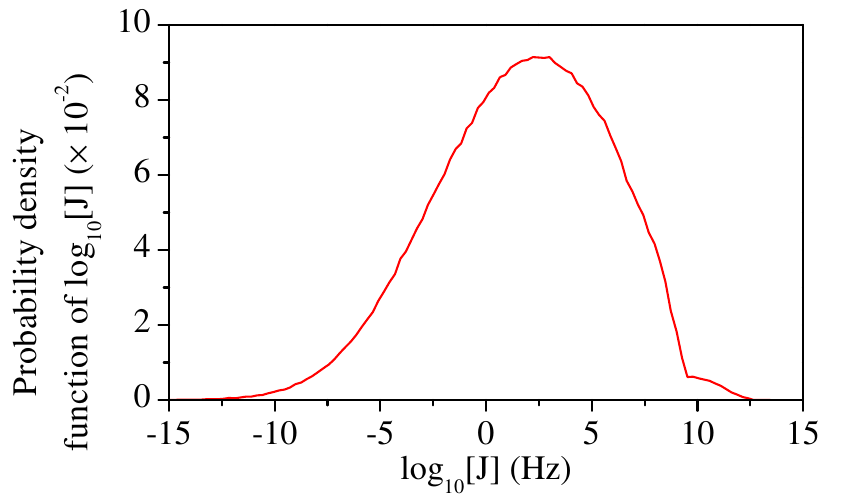}
\caption{Probability density function of $\log_{10}[\mathrm{J}]$ for a doping density of \SI{2E16}{P\per\cmc} in silicon. 
\label{fig:logJdistribution}}
\end{figure}

The extremely broad range of $J$ values resulting from these calculations is best displayed by calculating the probability density of $\log(J)$. The result is shown in Fig.~\ref{fig:logJdistribution} for a density of \SI{2E16}{\per\cmc} with $J$ coupling in dimers ranging from \SI{E-10}{\Hz} to \SI{E13}{\Hz}. The calculated $J$ distribution is in agreement with the results in Ref.~\onlinecite{Cullis1970}. Of this broad distribution of $J$, only the small number of dimers with $J \gg A = \SI{117}{\MHz}$ ($\log_{10}(J) > 8$) contribute to the center dimer line in the ESR experiment. The probability density function for these dimers is shown in Fig.~\ref{fig:Jdistribution}.

\begin{acknowledgments}
The authors thank Aman Jain, Xuedong Hu and John J. L. Morton for useful discussions. This work was supported by the NSF and EPSRC through the Materials World Network Program (DMR-1107606 and EP/I035536/1) and also by the ARO (W911NF-13-1-0179) and Princeton MRSEC (DMR-01420541).
\end{acknowledgments}


\begin{thebibliography}{40}%
\makeatletter
\providecommand \@ifxundefined [1]{%
 \@ifx{#1\undefined}
}%
\providecommand \@ifnum [1]{%
 \ifnum #1\expandafter \@firstoftwo
 \else \expandafter \@secondoftwo
 \fi
}%
\providecommand \@ifx [1]{%
 \ifx #1\expandafter \@firstoftwo
 \else \expandafter \@secondoftwo
 \fi
}%
\providecommand \natexlab [1]{#1}%
\providecommand \enquote  [1]{``#1''}%
\providecommand \bibnamefont  [1]{#1}%
\providecommand \bibfnamefont [1]{#1}%
\providecommand \citenamefont [1]{#1}%
\providecommand \href@noop [0]{\@secondoftwo}%
\providecommand \href [0]{\begingroup \@sanitize@url \@href}%
\providecommand \@href[1]{\@@startlink{#1}\@@href}%
\providecommand \@@href[1]{\endgroup#1\@@endlink}%
\providecommand \@sanitize@url [0]{\catcode `\\12\catcode `\$12\catcode
  `\&12\catcode `\#12\catcode `\^12\catcode `\_12\catcode `\%12\relax}%
\providecommand \@@startlink[1]{}%
\providecommand \@@endlink[0]{}%
\providecommand \url  [0]{\begingroup\@sanitize@url \@url }%
\providecommand \@url [1]{\endgroup\@href {#1}{\urlprefix }}%
\providecommand \urlprefix  [0]{URL }%
\providecommand \Eprint [0]{\href }%
\providecommand \doibase [0]{http://dx.doi.org/}%
\providecommand \selectlanguage [0]{\@gobble}%
\providecommand \bibinfo  [0]{\@secondoftwo}%
\providecommand \bibfield  [0]{\@secondoftwo}%
\providecommand \translation [1]{[#1]}%
\providecommand \BibitemOpen [0]{}%
\providecommand \bibitemStop [0]{}%
\providecommand \bibitemNoStop [0]{.\EOS\space}%
\providecommand \EOS [0]{\spacefactor3000\relax}%
\providecommand \BibitemShut  [1]{\csname bibitem#1\endcsname}%
\let\auto@bib@innerbib\@empty
\bibitem [{\citenamefont {Bulanov}\ \emph {et~al.}(2000)\citenamefont
  {Bulanov}, \citenamefont {Devyatych}, \citenamefont {Gusev}, \citenamefont
  {Sennikov}, \citenamefont {Pohl}, \citenamefont {Riemann}, \citenamefont
  {Schilling},\ and\ \citenamefont {Becker}}]{Bulanov2000}%
  \BibitemOpen
  \bibfield  {author} {\bibinfo {author} {\bibfnamefont {A.}~\bibnamefont
  {Bulanov}}, \bibinfo {author} {\bibfnamefont {G.}~\bibnamefont {Devyatych}},
  \bibinfo {author} {\bibfnamefont {A.}~\bibnamefont {Gusev}}, \bibinfo
  {author} {\bibfnamefont {P.}~\bibnamefont {Sennikov}}, \bibinfo {author}
  {\bibfnamefont {H.-J.}\ \bibnamefont {Pohl}}, \bibinfo {author}
  {\bibfnamefont {H.}~\bibnamefont {Riemann}}, \bibinfo {author} {\bibfnamefont
  {H.}~\bibnamefont {Schilling}}, \ and\ \bibinfo {author} {\bibfnamefont
  {P.}~\bibnamefont {Becker}},\ }\href {\doibase
  10.1002/1521-4079(200009)35:9<1023::AID-CRAT1023>3.0.CO;2-V} {\bibfield
  {journal} {\bibinfo  {journal} {Cryst. Res. Technol.}\ }\textbf {\bibinfo
  {volume} {35}},\ \bibinfo {pages} {1023} (\bibinfo {year}
  {2000})}\BibitemShut {NoStop}%
\bibitem [{\citenamefont {Itoh}\ \emph {et~al.}(2003)\citenamefont {Itoh},
  \citenamefont {Kato}, \citenamefont {Uemura}, \citenamefont {Kaliteevskii},
  \citenamefont {Godisov}, \citenamefont {Devyatych}, \citenamefont {Bulanov},
  \citenamefont {Gusev}, \citenamefont {Kovalev}, \citenamefont {Sennikov},
  \citenamefont {Pohl}, \citenamefont {Abrosimov},\ and\ \citenamefont
  {Riemann}}]{Itoh2003}%
  \BibitemOpen
  \bibfield  {author} {\bibinfo {author} {\bibfnamefont {K.~M.}\ \bibnamefont
  {Itoh}}, \bibinfo {author} {\bibfnamefont {J.}~\bibnamefont {Kato}}, \bibinfo
  {author} {\bibfnamefont {M.}~\bibnamefont {Uemura}}, \bibinfo {author}
  {\bibfnamefont {A.~K.}\ \bibnamefont {Kaliteevskii}}, \bibinfo {author}
  {\bibfnamefont {O.~N.}\ \bibnamefont {Godisov}}, \bibinfo {author}
  {\bibfnamefont {G.~G.}\ \bibnamefont {Devyatych}}, \bibinfo {author}
  {\bibfnamefont {A.~D.}\ \bibnamefont {Bulanov}}, \bibinfo {author}
  {\bibfnamefont {A.~V.}\ \bibnamefont {Gusev}}, \bibinfo {author}
  {\bibfnamefont {I.~D.}\ \bibnamefont {Kovalev}}, \bibinfo {author}
  {\bibfnamefont {P.~G.}\ \bibnamefont {Sennikov}}, \bibinfo {author}
  {\bibfnamefont {H.-J.}\ \bibnamefont {Pohl}}, \bibinfo {author}
  {\bibfnamefont {N.~V.}\ \bibnamefont {Abrosimov}}, \ and\ \bibinfo {author}
  {\bibfnamefont {H.}~\bibnamefont {Riemann}},\ }\href {\doibase
  10.1143/JJAP.42.6248} {\bibfield  {journal} {\bibinfo  {journal} {Jpn. J.
  Appl. Phys.}\ }\textbf {\bibinfo {volume} {42}},\ \bibinfo {pages} {6248}
  (\bibinfo {year} {2003})}\BibitemShut {NoStop}%
\bibitem [{\citenamefont {Ager~III}\ \emph {et~al.}(2005)\citenamefont
  {Ager~III}, \citenamefont {Beeman}, \citenamefont {Hansen}, \citenamefont
  {Haller}, \citenamefont {Sharp}, \citenamefont {Liao}, \citenamefont {Yang},
  \citenamefont {Thewalt},\ and\ \citenamefont {Riemann}}]{Ager2005}%
  \BibitemOpen
  \bibfield  {author} {\bibinfo {author} {\bibfnamefont {J.~W.}\ \bibnamefont
  {Ager~III}}, \bibinfo {author} {\bibfnamefont {J.~W.}\ \bibnamefont
  {Beeman}}, \bibinfo {author} {\bibfnamefont {W.~L.}\ \bibnamefont {Hansen}},
  \bibinfo {author} {\bibfnamefont {E.~E.}\ \bibnamefont {Haller}}, \bibinfo
  {author} {\bibfnamefont {I.~D.}\ \bibnamefont {Sharp}}, \bibinfo {author}
  {\bibfnamefont {C.}~\bibnamefont {Liao}}, \bibinfo {author} {\bibfnamefont
  {A.}~\bibnamefont {Yang}}, \bibinfo {author} {\bibfnamefont {M.~L.~W.}\
  \bibnamefont {Thewalt}}, \ and\ \bibinfo {author} {\bibfnamefont
  {H.}~\bibnamefont {Riemann}},\ }\href {\doibase 10.1149/1.1901674} {\bibfield
   {journal} {\bibinfo  {journal} {J. Electrochem. Soc.}\ }\textbf {\bibinfo
  {volume} {152}},\ \bibinfo {pages} {G448} (\bibinfo {year}
  {2005})}\BibitemShut {NoStop}%
\bibitem [{\citenamefont {Becker}(2003)}]{Becker2003}%
  \BibitemOpen
  \bibfield  {author} {\bibinfo {author} {\bibfnamefont {P.}~\bibnamefont
  {Becker}},\ }\href {\doibase 10.1088/0026-1394/40/6/008} {\bibfield
  {journal} {\bibinfo  {journal} {Metrologia}\ }\textbf {\bibinfo {volume}
  {40}},\ \bibinfo {pages} {366} (\bibinfo {year} {2003})}\BibitemShut
  {NoStop}%
\bibitem [{\citenamefont {Kane}(1998)}]{Kane1998}%
  \BibitemOpen
  \bibfield  {author} {\bibinfo {author} {\bibfnamefont {B.~E.}\ \bibnamefont
  {Kane}},\ }\href {\doibase 10.1038/30156} {\bibfield  {journal} {\bibinfo
  {journal} {Nature}\ }\textbf {\bibinfo {volume} {393}},\ \bibinfo {pages}
  {133} (\bibinfo {year} {1998})}\BibitemShut {NoStop}%
\bibitem [{\citenamefont {Cardona}\ and\ \citenamefont
  {Thewalt}(2005)}]{Cardona2005}%
  \BibitemOpen
  \bibfield  {author} {\bibinfo {author} {\bibfnamefont {M.}~\bibnamefont
  {Cardona}}\ and\ \bibinfo {author} {\bibfnamefont {M.~L.~W.}\ \bibnamefont
  {Thewalt}},\ }\href {\doibase 10.1103/RevModPhys.77.1173} {\bibfield
  {journal} {\bibinfo  {journal} {Rev. Mod. Phys.}\ }\textbf {\bibinfo {volume}
  {77}},\ \bibinfo {pages} {1173} (\bibinfo {year} {2005})}\BibitemShut
  {NoStop}%
\bibitem [{\citenamefont {Tezuka}\ \emph {et~al.}(2010)\citenamefont {Tezuka},
  \citenamefont {Stegner}, \citenamefont {Tyryshkin}, \citenamefont {Shankar},
  \citenamefont {Thewalt}, \citenamefont {Lyon}, \citenamefont {Itoh},\ and\
  \citenamefont {Brandt}}]{Tezuka2010}%
  \BibitemOpen
  \bibfield  {author} {\bibinfo {author} {\bibfnamefont {H.}~\bibnamefont
  {Tezuka}}, \bibinfo {author} {\bibfnamefont {A.~R.}\ \bibnamefont {Stegner}},
  \bibinfo {author} {\bibfnamefont {A.~M.}\ \bibnamefont {Tyryshkin}}, \bibinfo
  {author} {\bibfnamefont {S.}~\bibnamefont {Shankar}}, \bibinfo {author}
  {\bibfnamefont {M.~L.~W.}\ \bibnamefont {Thewalt}}, \bibinfo {author}
  {\bibfnamefont {S.~A.}\ \bibnamefont {Lyon}}, \bibinfo {author}
  {\bibfnamefont {K.~M.}\ \bibnamefont {Itoh}}, \ and\ \bibinfo {author}
  {\bibfnamefont {M.~S.}\ \bibnamefont {Brandt}},\ }\href {\doibase
  10.1103/PhysRevB.81.161203} {\bibfield  {journal} {\bibinfo  {journal} {Phys.
  Rev. B}\ }\textbf {\bibinfo {volume} {81}},\ \bibinfo {pages} {161203}
  (\bibinfo {year} {2010})}\BibitemShut {NoStop}%
\bibitem [{\citenamefont {Tyryshkin}\ \emph {et~al.}(2003)\citenamefont
  {Tyryshkin}, \citenamefont {Lyon}, \citenamefont {Astashkin},\ and\
  \citenamefont {Raitsimring}}]{Tyryshkin2003}%
  \BibitemOpen
  \bibfield  {author} {\bibinfo {author} {\bibfnamefont {A.~M.}\ \bibnamefont
  {Tyryshkin}}, \bibinfo {author} {\bibfnamefont {S.~A.}\ \bibnamefont {Lyon}},
  \bibinfo {author} {\bibfnamefont {A.~V.}\ \bibnamefont {Astashkin}}, \ and\
  \bibinfo {author} {\bibfnamefont {A.~M.}\ \bibnamefont {Raitsimring}},\
  }\href {\doibase 10.1103/PhysRevB.68.193207} {\bibfield  {journal} {\bibinfo
  {journal} {Phys. Rev. B}\ }\textbf {\bibinfo {volume} {68}},\ \bibinfo
  {pages} {193207} (\bibinfo {year} {2003})}\BibitemShut {NoStop}%
\bibitem [{\citenamefont {Morton}\ \emph {et~al.}(2011)\citenamefont {Morton},
  \citenamefont {McCamey}, \citenamefont {Eriksson},\ and\ \citenamefont
  {Lyon}}]{Morton2011}%
  \BibitemOpen
  \bibfield  {author} {\bibinfo {author} {\bibfnamefont {J.~J.~L.}\
  \bibnamefont {Morton}}, \bibinfo {author} {\bibfnamefont {D.~R.}\
  \bibnamefont {McCamey}}, \bibinfo {author} {\bibfnamefont {M.~A.}\
  \bibnamefont {Eriksson}}, \ and\ \bibinfo {author} {\bibfnamefont {S.~A.}\
  \bibnamefont {Lyon}},\ }\href {http://dx.doi.org/10.1038/nature10681}
  {\bibfield  {journal} {\bibinfo  {journal} {Nature}\ }\textbf {\bibinfo
  {volume} {479}},\ \bibinfo {pages} {345} (\bibinfo {year}
  {2011})}\BibitemShut {NoStop}%
\bibitem [{\citenamefont {Zwanenburg}\ \emph {et~al.}(2013)\citenamefont
  {Zwanenburg}, \citenamefont {Dzurak}, \citenamefont {Morello}, \citenamefont
  {Simmons}, \citenamefont {Hollenberg}, \citenamefont {Klimeck}, \citenamefont
  {Rogge}, \citenamefont {Coppersmith},\ and\ \citenamefont
  {Eriksson}}]{Zwanenburg2013}%
  \BibitemOpen
  \bibfield  {author} {\bibinfo {author} {\bibfnamefont {F.~A.}\ \bibnamefont
  {Zwanenburg}}, \bibinfo {author} {\bibfnamefont {A.~S.}\ \bibnamefont
  {Dzurak}}, \bibinfo {author} {\bibfnamefont {A.}~\bibnamefont {Morello}},
  \bibinfo {author} {\bibfnamefont {M.~Y.}\ \bibnamefont {Simmons}}, \bibinfo
  {author} {\bibfnamefont {L.~C.~L.}\ \bibnamefont {Hollenberg}}, \bibinfo
  {author} {\bibfnamefont {G.}~\bibnamefont {Klimeck}}, \bibinfo {author}
  {\bibfnamefont {S.}~\bibnamefont {Rogge}}, \bibinfo {author} {\bibfnamefont
  {S.~N.}\ \bibnamefont {Coppersmith}}, \ and\ \bibinfo {author} {\bibfnamefont
  {M.~A.}\ \bibnamefont {Eriksson}},\ }\href {\doibase
  10.1103/RevModPhys.85.961} {\bibfield  {journal} {\bibinfo  {journal} {Rev.
  Mod. Phys.}\ }\textbf {\bibinfo {volume} {85}},\ \bibinfo {pages} {961}
  (\bibinfo {year} {2013})}\BibitemShut {NoStop}%
\bibitem [{\citenamefont {Feher}(1959)}]{Feher1959}%
  \BibitemOpen
  \bibfield  {author} {\bibinfo {author} {\bibfnamefont {G.}~\bibnamefont
  {Feher}},\ }\href {\doibase 10.1103/PhysRev.114.1219} {\bibfield  {journal}
  {\bibinfo  {journal} {Phys. Rev.}\ }\textbf {\bibinfo {volume} {114}},\
  \bibinfo {pages} {1219} (\bibinfo {year} {1959})}\BibitemShut {NoStop}%
\bibitem [{\citenamefont {Feher}\ and\ \citenamefont
  {Gere}(1959)}]{FeherGere1959}%
  \BibitemOpen
  \bibfield  {author} {\bibinfo {author} {\bibfnamefont {G.}~\bibnamefont
  {Feher}}\ and\ \bibinfo {author} {\bibfnamefont {E.~A.}\ \bibnamefont
  {Gere}},\ }\href {\doibase 10.1103/PhysRev.114.1245} {\bibfield  {journal}
  {\bibinfo  {journal} {Phys. Rev.}\ }\textbf {\bibinfo {volume} {114}},\
  \bibinfo {pages} {1245} (\bibinfo {year} {1959})}\BibitemShut {NoStop}%
\bibitem [{\citenamefont {Wilson}\ and\ \citenamefont
  {Feher}(1961)}]{Wilson1961}%
  \BibitemOpen
  \bibfield  {author} {\bibinfo {author} {\bibfnamefont {D.~K.}\ \bibnamefont
  {Wilson}}\ and\ \bibinfo {author} {\bibfnamefont {G.}~\bibnamefont {Feher}},\
  }\href {\doibase 10.1103/PhysRev.124.1068} {\bibfield  {journal} {\bibinfo
  {journal} {Phys. Rev.}\ }\textbf {\bibinfo {volume} {124}},\ \bibinfo {pages}
  {1068} (\bibinfo {year} {1961})}\BibitemShut {NoStop}%
\bibitem [{\citenamefont {Abe}\ \emph {et~al.}(2010)\citenamefont {Abe},
  \citenamefont {Tyryshkin}, \citenamefont {Tojo}, \citenamefont {Morton},
  \citenamefont {Witzel}, \citenamefont {Fujimoto}, \citenamefont {Ager},
  \citenamefont {Haller}, \citenamefont {Isoya}, \citenamefont {Lyon},
  \citenamefont {Thewalt},\ and\ \citenamefont {Itoh}}]{Abe2010}%
  \BibitemOpen
  \bibfield  {author} {\bibinfo {author} {\bibfnamefont {E.}~\bibnamefont
  {Abe}}, \bibinfo {author} {\bibfnamefont {A.~M.}\ \bibnamefont {Tyryshkin}},
  \bibinfo {author} {\bibfnamefont {S.}~\bibnamefont {Tojo}}, \bibinfo {author}
  {\bibfnamefont {J.~J.~L.}\ \bibnamefont {Morton}}, \bibinfo {author}
  {\bibfnamefont {W.~M.}\ \bibnamefont {Witzel}}, \bibinfo {author}
  {\bibfnamefont {A.}~\bibnamefont {Fujimoto}}, \bibinfo {author}
  {\bibfnamefont {J.~W.}\ \bibnamefont {Ager}}, \bibinfo {author}
  {\bibfnamefont {E.~E.}\ \bibnamefont {Haller}}, \bibinfo {author}
  {\bibfnamefont {J.}~\bibnamefont {Isoya}}, \bibinfo {author} {\bibfnamefont
  {S.~A.}\ \bibnamefont {Lyon}}, \bibinfo {author} {\bibfnamefont {M.~L.~W.}\
  \bibnamefont {Thewalt}}, \ and\ \bibinfo {author} {\bibfnamefont {K.~M.}\
  \bibnamefont {Itoh}},\ }\href {\doibase 10.1103/PhysRevB.82.121201}
  {\bibfield  {journal} {\bibinfo  {journal} {Phys. Rev. B}\ }\textbf {\bibinfo
  {volume} {82}},\ \bibinfo {pages} {121201} (\bibinfo {year}
  {2010})}\BibitemShut {NoStop}%
\bibitem [{\citenamefont {Feher}\ \emph {et~al.}(1955)\citenamefont {Feher},
  \citenamefont {Fletcher},\ and\ \citenamefont {Gere}}]{Feher1955}%
  \BibitemOpen
  \bibfield  {author} {\bibinfo {author} {\bibfnamefont {G.}~\bibnamefont
  {Feher}}, \bibinfo {author} {\bibfnamefont {R.}~\bibnamefont {Fletcher}}, \
  and\ \bibinfo {author} {\bibfnamefont {E.}~\bibnamefont {Gere}},\ }\href
  {\doibase 10.1103/PhysRev.100.1784.2} {\bibfield  {journal} {\bibinfo
  {journal} {Phys. Rev.}\ }\textbf {\bibinfo {volume} {100}},\ \bibinfo {pages}
  {1784} (\bibinfo {year} {1955})}\BibitemShut {NoStop}%
\bibitem [{\citenamefont {Slichter}(1955)}]{Slichter1955}%
  \BibitemOpen
  \bibfield  {author} {\bibinfo {author} {\bibfnamefont {C.}~\bibnamefont
  {Slichter}},\ }\href {\doibase 10.1103/PhysRev.99.479} {\bibfield  {journal}
  {\bibinfo  {journal} {Phys. Rev.}\ }\textbf {\bibinfo {volume} {99}},\
  \bibinfo {pages} {479} (\bibinfo {year} {1955})}\BibitemShut {NoStop}%
\bibitem [{\citenamefont {Cullis}\ and\ \citenamefont
  {Marko}(1970)}]{Cullis1970}%
  \BibitemOpen
  \bibfield  {author} {\bibinfo {author} {\bibfnamefont {P.}~\bibnamefont
  {Cullis}}\ and\ \bibinfo {author} {\bibfnamefont {J.}~\bibnamefont {Marko}},\
  }\href {\doibase 10.1103/PhysRevB.1.632} {\bibfield  {journal} {\bibinfo
  {journal} {Phys. Rev. B}\ }\textbf {\bibinfo {volume} {1}},\ \bibinfo {pages}
  {632} (\bibinfo {year} {1970})}\BibitemShut {NoStop}%
\bibitem [{\citenamefont {Vrijen}\ \emph {et~al.}(2000)\citenamefont {Vrijen},
  \citenamefont {Yablonovitch}, \citenamefont {Wang}, \citenamefont {Jiang},
  \citenamefont {Balandin}, \citenamefont {Roychowdhury}, \citenamefont {Mor},\
  and\ \citenamefont {DiVincenzo}}]{Vrijen2000}%
  \BibitemOpen
  \bibfield  {author} {\bibinfo {author} {\bibfnamefont {R.}~\bibnamefont
  {Vrijen}}, \bibinfo {author} {\bibfnamefont {E.}~\bibnamefont
  {Yablonovitch}}, \bibinfo {author} {\bibfnamefont {K.}~\bibnamefont {Wang}},
  \bibinfo {author} {\bibfnamefont {H.~W.}\ \bibnamefont {Jiang}}, \bibinfo
  {author} {\bibfnamefont {A.}~\bibnamefont {Balandin}}, \bibinfo {author}
  {\bibfnamefont {V.}~\bibnamefont {Roychowdhury}}, \bibinfo {author}
  {\bibfnamefont {T.}~\bibnamefont {Mor}}, \ and\ \bibinfo {author}
  {\bibfnamefont {D.~P.}\ \bibnamefont {DiVincenzo}},\ }\href {\doibase
  10.1103/PhysRevA.62.012306} {\bibfield  {journal} {\bibinfo  {journal} {Phys.
  Rev. A}\ }\textbf {\bibinfo {volume} {62}},\ \bibinfo {pages} {012306}
  (\bibinfo {year} {2000})}\BibitemShut {NoStop}%
\bibitem [{\citenamefont {Culcer}\ \emph {et~al.}(2009)\citenamefont {Culcer},
  \citenamefont {Hu},\ and\ \citenamefont {Das~Sarma}}]{Culcer2009}%
  \BibitemOpen
  \bibfield  {author} {\bibinfo {author} {\bibfnamefont {D.}~\bibnamefont
  {Culcer}}, \bibinfo {author} {\bibfnamefont {X.}~\bibnamefont {Hu}}, \ and\
  \bibinfo {author} {\bibfnamefont {S.}~\bibnamefont {Das~Sarma}},\ }\href
  {\doibase http://dx.doi.org/10.1063/1.3194778} {\bibfield  {journal}
  {\bibinfo  {journal} {Applied Physics Letters}\ }\textbf {\bibinfo {volume}
  {95}},\ \bibinfo {eid} {073102} (\bibinfo {year} {2009})}\BibitemShut
  {NoStop}%
\bibitem [{\citenamefont {Borhani}\ and\ \citenamefont
  {Hu}(2010)}]{XuedongHu2010}%
  \BibitemOpen
  \bibfield  {author} {\bibinfo {author} {\bibfnamefont {M.}~\bibnamefont
  {Borhani}}\ and\ \bibinfo {author} {\bibfnamefont {X.}~\bibnamefont {Hu}},\
  }\href@noop {} {\bibfield  {journal} {\bibinfo  {journal} {Phys. Rev. B}\
  }\textbf {\bibinfo {volume} {82}},\ \bibinfo {pages} {241302(R)} (\bibinfo
  {year} {2010})}\BibitemShut {NoStop}%
\bibitem [{\citenamefont {Tyryshkin}\ \emph {et~al.}(2012)\citenamefont
  {Tyryshkin}, \citenamefont {Tojo}, \citenamefont {Morton}, \citenamefont
  {Riemann}, \citenamefont {Abrosimov}, \citenamefont {Becker}, \citenamefont
  {Pohl}, \citenamefont {Schenkel}, \citenamefont {Thewalt}, \citenamefont
  {Itoh},\ and\ \citenamefont {Lyon}}]{Tyryshkin2011}%
  \BibitemOpen
  \bibfield  {author} {\bibinfo {author} {\bibfnamefont {A.~M.}\ \bibnamefont
  {Tyryshkin}}, \bibinfo {author} {\bibfnamefont {S.}~\bibnamefont {Tojo}},
  \bibinfo {author} {\bibfnamefont {J.~J.~L.}\ \bibnamefont {Morton}}, \bibinfo
  {author} {\bibfnamefont {H.}~\bibnamefont {Riemann}}, \bibinfo {author}
  {\bibfnamefont {N.~V.}\ \bibnamefont {Abrosimov}}, \bibinfo {author}
  {\bibfnamefont {P.}~\bibnamefont {Becker}}, \bibinfo {author} {\bibfnamefont
  {H.-J.}\ \bibnamefont {Pohl}}, \bibinfo {author} {\bibfnamefont
  {T.}~\bibnamefont {Schenkel}}, \bibinfo {author} {\bibfnamefont {M.~L.~W.}\
  \bibnamefont {Thewalt}}, \bibinfo {author} {\bibfnamefont {K.~M.}\
  \bibnamefont {Itoh}}, \ and\ \bibinfo {author} {\bibfnamefont {S.~A.}\
  \bibnamefont {Lyon}},\ }\href {\doibase 10.1038/nmat3182} {\bibfield
  {journal} {\bibinfo  {journal} {Nat. Mater.}\ }\textbf {\bibinfo {volume}
  {11}},\ \bibinfo {pages} {143} (\bibinfo {year} {2012})}\BibitemShut
  {NoStop}%
\bibitem [{\citenamefont {Dehollain}\ \emph {et~al.}(2014)\citenamefont
  {Dehollain}, \citenamefont {Muhonen}, \citenamefont {Tan}, \citenamefont
  {Saraiva}, \citenamefont {Jamieson}, \citenamefont {Dzurak},\ and\
  \citenamefont {Morello}}]{Dehollain2014}%
  \BibitemOpen
  \bibfield  {author} {\bibinfo {author} {\bibfnamefont {J.~P.}\ \bibnamefont
  {Dehollain}}, \bibinfo {author} {\bibfnamefont {J.~T.}\ \bibnamefont
  {Muhonen}}, \bibinfo {author} {\bibfnamefont {K.~Y.}\ \bibnamefont {Tan}},
  \bibinfo {author} {\bibfnamefont {A.}~\bibnamefont {Saraiva}}, \bibinfo
  {author} {\bibfnamefont {D.~N.}\ \bibnamefont {Jamieson}}, \bibinfo {author}
  {\bibfnamefont {A.~S.}\ \bibnamefont {Dzurak}}, \ and\ \bibinfo {author}
  {\bibfnamefont {A.}~\bibnamefont {Morello}},\ }\href {\doibase
  10.1103/PhysRevLett.112.236801} {\bibfield  {journal} {\bibinfo  {journal}
  {Phys. Rev. Lett.}\ }\textbf {\bibinfo {volume} {112}},\ \bibinfo {pages}
  {236801} (\bibinfo {year} {2014})}\BibitemShut {NoStop}%
\bibitem [{\citenamefont {Hahn}(1950)}]{Hahn1950}%
  \BibitemOpen
  \bibfield  {author} {\bibinfo {author} {\bibfnamefont {E.~L.}\ \bibnamefont
  {Hahn}},\ }\href {\doibase 10.1103/PhysRev.80.580} {\bibfield  {journal}
  {\bibinfo  {journal} {Phys. Rev.}\ }\textbf {\bibinfo {volume} {80}},\
  \bibinfo {pages} {580} (\bibinfo {year} {1950})}\BibitemShut {NoStop}%
\bibitem [{\citenamefont {Morton}\ \emph {et~al.}(2005)\citenamefont {Morton},
  \citenamefont {Tyryshkin}, \citenamefont {Ardavan}, \citenamefont
  {Porfyrakis}, \citenamefont {Lyon},\ and\ \citenamefont
  {Briggs}}]{Morton2005}%
  \BibitemOpen
  \bibfield  {author} {\bibinfo {author} {\bibfnamefont {J.~J.~L.}\
  \bibnamefont {Morton}}, \bibinfo {author} {\bibfnamefont {A.~M.}\
  \bibnamefont {Tyryshkin}}, \bibinfo {author} {\bibfnamefont {A.}~\bibnamefont
  {Ardavan}}, \bibinfo {author} {\bibfnamefont {K.}~\bibnamefont {Porfyrakis}},
  \bibinfo {author} {\bibfnamefont {S.~A.}\ \bibnamefont {Lyon}}, \ and\
  \bibinfo {author} {\bibfnamefont {G.~A.~D.}\ \bibnamefont {Briggs}},\ }\href
  {\doibase 10.1063/1.1888585} {\bibfield  {journal} {\bibinfo  {journal} {J.
  Chem. Phys.}\ }\textbf {\bibinfo {volume} {122}},\ \bibinfo {pages} {174504}
  (\bibinfo {year} {2005})}\BibitemShut {NoStop}%
\bibitem [{\citenamefont {Schweiger}\ and\ \citenamefont
  {Jeschke}(2001)}]{Schweiger2001}%
  \BibitemOpen
  \bibfield  {author} {\bibinfo {author} {\bibfnamefont {A.}~\bibnamefont
  {Schweiger}}\ and\ \bibinfo {author} {\bibfnamefont {G.}~\bibnamefont
  {Jeschke}},\ }\href@noop {} {\emph {\bibinfo {title} {Principles of pulse
  electron paramagnetic resonance}}}\ (\bibinfo  {publisher} {Oxford University
  Press},\ \bibinfo {year} {2001})\BibitemShut {NoStop}%
\bibitem [{\citenamefont {Klauder}\ and\ \citenamefont
  {Anderson}(1962)}]{Klauder1962}%
  \BibitemOpen
  \bibfield  {author} {\bibinfo {author} {\bibfnamefont {J.~R.}\ \bibnamefont
  {Klauder}}\ and\ \bibinfo {author} {\bibfnamefont {P.~W.}\ \bibnamefont
  {Anderson}},\ }\href {\doibase 10.1103/PhysRev.125.912} {\bibfield  {journal}
  {\bibinfo  {journal} {Phys. Rev.}\ }\textbf {\bibinfo {volume} {125}},\
  \bibinfo {pages} {912} (\bibinfo {year} {1962})}\BibitemShut {NoStop}%
\bibitem [{\citenamefont {Mims}(1968)}]{Mims1968}%
  \BibitemOpen
  \bibfield  {author} {\bibinfo {author} {\bibfnamefont {W.~B.}\ \bibnamefont
  {Mims}},\ }\href {\doibase 10.1103/PhysRev.168.370} {\bibfield  {journal}
  {\bibinfo  {journal} {Phys. Rev.}\ }\textbf {\bibinfo {volume} {168}},\
  \bibinfo {pages} {370} (\bibinfo {year} {1968})}\BibitemShut {NoStop}%
\bibitem [{\citenamefont {Stoll}\ and\ \citenamefont
  {Schweiger}(2006)}]{Stoll2006}%
  \BibitemOpen
  \bibfield  {author} {\bibinfo {author} {\bibfnamefont {S.}~\bibnamefont
  {Stoll}}\ and\ \bibinfo {author} {\bibfnamefont {A.}~\bibnamefont
  {Schweiger}},\ }\href {\doibase DOI: 10.1016/j.jmr.2005.08.013} {\bibfield
  {journal} {\bibinfo  {journal} {J. Mag. Res.}\ }\textbf {\bibinfo {volume}
  {178}},\ \bibinfo {pages} {42 } (\bibinfo {year} {2006})}\BibitemShut
  {NoStop}%
\bibitem [{\citenamefont {Tyryshkin}\ \emph {et~al.}(2006)\citenamefont
  {Tyryshkin}, \citenamefont {Morton}, \citenamefont {Benjamin}, \citenamefont
  {Ardavan}, \citenamefont {Briggs}, \citenamefont {Ager},\ and\ \citenamefont
  {Lyon}}]{Tyryshkin2006b}%
  \BibitemOpen
  \bibfield  {author} {\bibinfo {author} {\bibfnamefont {A.~M.}\ \bibnamefont
  {Tyryshkin}}, \bibinfo {author} {\bibfnamefont {J.~J.~L.}\ \bibnamefont
  {Morton}}, \bibinfo {author} {\bibfnamefont {S.~C.}\ \bibnamefont
  {Benjamin}}, \bibinfo {author} {\bibfnamefont {A.}~\bibnamefont {Ardavan}},
  \bibinfo {author} {\bibfnamefont {G.~A.~D.}\ \bibnamefont {Briggs}}, \bibinfo
  {author} {\bibfnamefont {J.~W.}\ \bibnamefont {Ager}}, \ and\ \bibinfo
  {author} {\bibfnamefont {S.~A.}\ \bibnamefont {Lyon}},\ }\href {\doibase
  10.1088/0953-8984/18/21/S06} {\bibfield  {journal} {\bibinfo  {journal} {J.
  Phys. Cond. Mat.}\ }\textbf {\bibinfo {volume} {18}},\ \bibinfo {pages}
  {S783} (\bibinfo {year} {2006})}\BibitemShut {NoStop}%
\bibitem [{\citenamefont {Biercuk}\ \emph {et~al.}(2009)\citenamefont
  {Biercuk}, \citenamefont {Uys}, \citenamefont {VanDevender}, \citenamefont
  {Shiga}, \citenamefont {Itano},\ and\ \citenamefont
  {Bollinger}}]{Biercuk2009}%
  \BibitemOpen
  \bibfield  {author} {\bibinfo {author} {\bibfnamefont {M.~J.}\ \bibnamefont
  {Biercuk}}, \bibinfo {author} {\bibfnamefont {H.}~\bibnamefont {Uys}},
  \bibinfo {author} {\bibfnamefont {A.~P.}\ \bibnamefont {VanDevender}},
  \bibinfo {author} {\bibfnamefont {N.}~\bibnamefont {Shiga}}, \bibinfo
  {author} {\bibfnamefont {W.~M.}\ \bibnamefont {Itano}}, \ and\ \bibinfo
  {author} {\bibfnamefont {J.~J.}\ \bibnamefont {Bollinger}},\ }\href
  {http://dx.doi.org/10.1038/nature07951} {\bibfield  {journal} {\bibinfo
  {journal} {Nature}\ }\textbf {\bibinfo {volume} {458}},\ \bibinfo {pages}
  {996} (\bibinfo {year} {2009})}\BibitemShut {NoStop}%
\bibitem [{\citenamefont {Saeedi}\ \emph {et~al.}(2013)\citenamefont {Saeedi},
  \citenamefont {Simmons}, \citenamefont {Salvail}, \citenamefont {Dluhy},
  \citenamefont {Riemann}, \citenamefont {Abrosimov}, \citenamefont {Becker},
  \citenamefont {Pohl}, \citenamefont {Morton},\ and\ \citenamefont
  {Thewalt}}]{Saeedi2013}%
  \BibitemOpen
  \bibfield  {author} {\bibinfo {author} {\bibfnamefont {K.}~\bibnamefont
  {Saeedi}}, \bibinfo {author} {\bibfnamefont {S.}~\bibnamefont {Simmons}},
  \bibinfo {author} {\bibfnamefont {J.~Z.}\ \bibnamefont {Salvail}}, \bibinfo
  {author} {\bibfnamefont {P.}~\bibnamefont {Dluhy}}, \bibinfo {author}
  {\bibfnamefont {H.}~\bibnamefont {Riemann}}, \bibinfo {author} {\bibfnamefont
  {N.~V.}\ \bibnamefont {Abrosimov}}, \bibinfo {author} {\bibfnamefont
  {P.}~\bibnamefont {Becker}}, \bibinfo {author} {\bibfnamefont {H.-J.}\
  \bibnamefont {Pohl}}, \bibinfo {author} {\bibfnamefont {J.~J.~L.}\
  \bibnamefont {Morton}}, \ and\ \bibinfo {author} {\bibfnamefont {M.~L.~W.}\
  \bibnamefont {Thewalt}},\ }\href {\doibase 10.1126/science.1239584}
  {\bibfield  {journal} {\bibinfo  {journal} {Science}\ }\textbf {\bibinfo
  {volume} {342}},\ \bibinfo {pages} {830} (\bibinfo {year}
  {2013})}\BibitemShut {NoStop}%
\bibitem [{\citenamefont {Castner}(1967)}]{Castner1967}%
  \BibitemOpen
  \bibfield  {author} {\bibinfo {author} {\bibfnamefont {T.~G.}\ \bibnamefont
  {Castner}},\ }\href {\doibase 10.1103/PhysRev.155.816} {\bibfield  {journal}
  {\bibinfo  {journal} {Phys. Rev.}\ }\textbf {\bibinfo {volume} {155}},\
  \bibinfo {pages} {816} (\bibinfo {year} {1967})}\BibitemShut {NoStop}%
\bibitem [{\citenamefont {Yudanov}\ \emph {et~al.}(1969)\citenamefont
  {Yudanov}, \citenamefont {Salikhov}, \citenamefont {Zhidomirov},\ and\
  \citenamefont {Tsvetkov}}]{Yudanov1969}%
  \BibitemOpen
  \bibfield  {author} {\bibinfo {author} {\bibfnamefont {V.~F.}\ \bibnamefont
  {Yudanov}}, \bibinfo {author} {\bibfnamefont {K.~M.}\ \bibnamefont
  {Salikhov}}, \bibinfo {author} {\bibfnamefont {G.~M.}\ \bibnamefont
  {Zhidomirov}}, \ and\ \bibinfo {author} {\bibfnamefont {Y.~D.}\ \bibnamefont
  {Tsvetkov}},\ }\href@noop {} {\bibfield  {journal} {\bibinfo  {journal}
  {Teor. Eksp. Khim.}\ }\textbf {\bibinfo {volume} {5}},\ \bibinfo {pages}
  {663} (\bibinfo {year} {1969})}\BibitemShut {NoStop}%
\bibitem [{\citenamefont {Milov}\ and\ \citenamefont
  {Tsvetkov}(1986)}]{Milov1986}%
  \BibitemOpen
  \bibfield  {author} {\bibinfo {author} {\bibfnamefont {A.~D.}\ \bibnamefont
  {Milov}}\ and\ \bibinfo {author} {\bibfnamefont {Y.~D.}\ \bibnamefont
  {Tsvetkov}},\ }\href@noop {} {\bibfield  {journal} {\bibinfo  {journal}
  {Dokl. Akad. Nauk SSSR}\ }\textbf {\bibinfo {volume} {288}},\ \bibinfo
  {pages} {924} (\bibinfo {year} {1986})}\BibitemShut {NoStop}%
\bibitem [{\citenamefont {Schenkel}\ \emph {et~al.}(2006)\citenamefont
  {Schenkel}, \citenamefont {Liddle}, \citenamefont {Persaud}, \citenamefont
  {Tyryshkin}, \citenamefont {Lyon}, \citenamefont {de~Sousa}, \citenamefont
  {Whaley}, \citenamefont {Bokor}, \citenamefont {Shangkuan},\ and\
  \citenamefont {Chakarov}}]{Schenkel2006}%
  \BibitemOpen
  \bibfield  {author} {\bibinfo {author} {\bibfnamefont {T.}~\bibnamefont
  {Schenkel}}, \bibinfo {author} {\bibfnamefont {J.~A.}\ \bibnamefont
  {Liddle}}, \bibinfo {author} {\bibfnamefont {A.}~\bibnamefont {Persaud}},
  \bibinfo {author} {\bibfnamefont {A.~M.}\ \bibnamefont {Tyryshkin}}, \bibinfo
  {author} {\bibfnamefont {S.~A.}\ \bibnamefont {Lyon}}, \bibinfo {author}
  {\bibfnamefont {R.}~\bibnamefont {de~Sousa}}, \bibinfo {author}
  {\bibfnamefont {K.~B.}\ \bibnamefont {Whaley}}, \bibinfo {author}
  {\bibfnamefont {J.}~\bibnamefont {Bokor}}, \bibinfo {author} {\bibfnamefont
  {J.}~\bibnamefont {Shangkuan}}, \ and\ \bibinfo {author} {\bibfnamefont
  {I.}~\bibnamefont {Chakarov}},\ }\href {\doibase 10.1063/1.2182068}
  {\bibfield  {journal} {\bibinfo  {journal} {Appl. Phys. Lett.}\ }\textbf
  {\bibinfo {volume} {88}},\ \bibinfo {eid} {112101} (\bibinfo {year}
  {2006})}\BibitemShut {NoStop}%
\bibitem [{\citenamefont {Andres}\ \emph {et~al.}(1981)\citenamefont {Andres},
  \citenamefont {Bhatt}, \citenamefont {Goalwin}, \citenamefont {Rice},\ and\
  \citenamefont {Walstedt}}]{Andres1981}%
  \BibitemOpen
  \bibfield  {author} {\bibinfo {author} {\bibfnamefont {K.}~\bibnamefont
  {Andres}}, \bibinfo {author} {\bibfnamefont {R.}~\bibnamefont {Bhatt}},
  \bibinfo {author} {\bibfnamefont {P.}~\bibnamefont {Goalwin}}, \bibinfo
  {author} {\bibfnamefont {T.}~\bibnamefont {Rice}}, \ and\ \bibinfo {author}
  {\bibfnamefont {R.}~\bibnamefont {Walstedt}},\ }\href {\doibase
  10.1103/PhysRevB.24.244} {\bibfield  {journal} {\bibinfo  {journal} {Phys.
  Rev. B}\ }\textbf {\bibinfo {volume} {24}},\ \bibinfo {pages} {244} (\bibinfo
  {year} {1981})}\BibitemShut {NoStop}%
\bibitem [{\citenamefont {New}\ and\ \citenamefont {Castner}(1984)}]{New1984}%
  \BibitemOpen
  \bibfield  {author} {\bibinfo {author} {\bibfnamefont {D.}~\bibnamefont
  {New}}\ and\ \bibinfo {author} {\bibfnamefont {T.~G.}\ \bibnamefont
  {Castner}},\ }\href {\doibase 10.1103/PhysRevB.29.2077} {\bibfield  {journal}
  {\bibinfo  {journal} {Phys. Rev. B}\ }\textbf {\bibinfo {volume} {29}},\
  \bibinfo {pages} {2077} (\bibinfo {year} {1984})}\BibitemShut {NoStop}%
\bibitem [{\citenamefont {New}(1985)}]{New1985}%
  \BibitemOpen
  \bibfield  {author} {\bibinfo {author} {\bibfnamefont {D.}~\bibnamefont
  {New}},\ }\href {\doibase 10.1103/PhysRevB.32.2419} {\bibfield  {journal}
  {\bibinfo  {journal} {Phys. Rev. B}\ }\textbf {\bibinfo {volume} {32}},\
  \bibinfo {pages} {2419} (\bibinfo {year} {1985})}\BibitemShut {NoStop}%
\bibitem [{\citenamefont {Ashcroft}\ and\ \citenamefont
  {Mermin}(1976)}]{AM1976}%
  \BibitemOpen
  \bibfield  {author} {\bibinfo {author} {\bibfnamefont {N.~W.}\ \bibnamefont
  {Ashcroft}}\ and\ \bibinfo {author} {\bibfnamefont {N.~D.}\ \bibnamefont
  {Mermin}},\ }\href@noop {} {\emph {\bibinfo {title} {Solid state physics}}}\
  (\bibinfo  {publisher} {Harcourt},\ \bibinfo {year} {1976})\BibitemShut
  {NoStop}%
\bibitem [{\citenamefont {Koiller}\ \emph {et~al.}(2001)\citenamefont
  {Koiller}, \citenamefont {Hu},\ and\ \citenamefont {{Das
  Sarma}}}]{Koiller2001}%
  \BibitemOpen
  \bibfield  {author} {\bibinfo {author} {\bibfnamefont {B.}~\bibnamefont
  {Koiller}}, \bibinfo {author} {\bibfnamefont {X.}~\bibnamefont {Hu}}, \ and\
  \bibinfo {author} {\bibfnamefont {S.}~\bibnamefont {{Das Sarma}}},\ }\href
  {\doibase 10.1103/PhysRevLett.88.027903} {\bibfield  {journal} {\bibinfo
  {journal} {Phys. Rev. Lett.}\ }\textbf {\bibinfo {volume} {88}},\ \bibinfo
  {pages} {027903} (\bibinfo {year} {2001})}\BibitemShut {NoStop}%
\end{thebibliography}
\end{document}